\documentclass[]{pasj02} 
\usepackage{graphicx}
\usepackage{tablefootnote}
\usepackage{amsmath}
\usepackage{threeparttable}
\usepackage{natbib}
\usepackage{multirow}
\usepackage[switch,mathlines]{lineno} 
\usepackage{caption}
\jyear{2024}
\Received{}
\Accepted{}

\begin{document} 

\title{Dynamics of the intermediate-mass-element ejecta in the Supernova Remnant Cassiopeia~A studied with XRISM}

\author{
 Shunsuke \textsc{Suzuki},\altaffilmark{1, 2}\orcid{0009-0008-1853-6379} \email{suzukishunsuke@phys.aoyama.ac.jp}
 Haruto \textsc{Sonoda},\altaffilmark{3, 2}\orcid{0009-0006-0015-4132} \email{sonoda-haruto397@g.ecc.u-tokyo.ac.jp}
 Yusuke \textsc{Sakai},\altaffilmark{4}\orcid{0000-0002-5809-3516}
 Yuken \textsc{Ohshiro},\altaffilmark{3, 2}\orcid{0009-0002-4783-3395}
 Shinya \textsc{Yamada},\altaffilmark{4}\orcid{0000-0003-4808-893X}
 Manan \textsc{Agarwal},\altaffilmark{5}\orcid{0000-0001-6965-8642}
 Satoru \textsc{Katsuda},\altaffilmark{6}\orcid{0000-0002-1104-7205}
 and 
 Hiroya \textsc{Yamaguchi}\altaffilmark{2, 3, 1}\orcid{0000-0002-5092-6085} \email{yamaguchi@astro.isas.jaxa.jp}
}
\altaffiltext{1}{Department of Science and Engineering, Graduate School of Science and Engineering, Aoyama Gakuin University, 5-10-1, Fuchinobe, Sagamihara 252-5258, Japan}
\altaffiltext{2}{Institute of Space and Astronautical Science (ISAS), Japan Aerospace Exploration Agency (JAXA), 3-1-1 Yoshinodai, Chuo-ku,
Sagamihara, Kanagawa 252-5210, Japan}
\altaffiltext{3}{Department of Physics, Graduate School of Science, The University of Tokyo, 7-3-1 Hongo, Bunkyo-ku, Tokyo 113-0033, Japan}
\altaffiltext{4}{Department of Physics, Rikkyo University, Toshima-Ku, Tokyo, 171-8501, Japan}
\altaffiltext{5}{Anton Pannekoek Institute/GRAPPA, University of Amsterdam, Science Park 904, 1098 XH Amsterdam, The Netherlands}
\altaffiltext{6}{Graduate School of Science and Engineering, Saitama University, 255 Shimo-Ohkubo, Sakura, Saitama 338-8570, Japan}




\maketitle

\begin{abstract}
Supernova remnants (SNRs) provide crucial information of yet poorly understood mechanism of supernova explosion. Here we present XRISM high-resolution spectroscopy of the intermediate-mass-element (IME) ejecta in the SNR Cas~A to determine their velocity distribution and thermal properties. 
The XRISM/Resolve spectrum in the 1.75--2.95~keV band extracted from each $1' \times 1'$ region in the southeast and northwest rims is fitted with a model consisting of two-component plasmas in non-equilibrium ionization with different radial velocities and ionization timescales. It is found that the more highly ionized component has a larger radial velocity, suggesting that this component is distributed in the outer layer and thus has been heated by the SNR reverse shock earlier. 
We also perform proper motion measurement of the highly ionized component (represented by the Si~\emissiontype{XIV} Ly$\alpha$ emission), using archival Chandra data, to reconstruct the three-dimensional velocity of the outermost IME ejecta. 
The pre-shock (free expansion) velocity of these ejecta is estimated to range from 2400 to 7100~km\,s$^{-1}$, based on the thermal properties and the bulk velocity of the shocked ejecta.
These velocities are consistent with theoretical predictions for a Type IIb supernova, in which the progenitor's hydrogen envelope is largely stripped before the explosion.
Furthermore, we find a substantial asymmetry in the distribution of the free expansion velocities, with the highest value toward the direction opposite to the proper motion of the neutron star (NS). 
This indicates the physical association between the asymmetric supernova explosion and NS kick.

\end{abstract}


\section{Introduction}

Supernovae (SNe) play a crucial role in the chemical and dynamical evolution of galaxies by releasing heavy elements synthesized in the progenitor and injecting huge energy into the ambient interstellar medium (ISM). 
Core-collapse SNe are also important as the origin of compact objects such as neutron stars (NSs) and black holes. 
However, their explosion mechanism is still poorly understood. 
Theoretical studies suggest that asymmetric effects play a key role in explosion \citep[e.g.,][]{1995ApJ...450..830B,2012ARNPS..62..407J,2017ApJ...837...84J,2013A&A...552A.126W,2017ApJ...842...13W}. 
Observations of supernova remnants (SNRs) provide essential clues to such asymmetric explosion effects, since the dynamics of SNRs can be investigated in detail \citep[e.g.,][]{DeLaney_2010,Milisavljevic_2013,Law_2020,2022ApJ...932..117L,2024ApJ...965L..27M}.

Cassiopeia~A (Cas~A) is the X-ray brightest SNR in our Galaxy, offering an ideal site to study the explosion mechanism of core-collapse SNe. 
Cas~A is believed to have originated from a Type IIb SN \citep{2008Sci...320.1195K,2011ApJ...732....3R} that occurred in the late 17th century \citep{2001AJ....122..297T}. 
Its progenitor is thought to be a red supergiant with the zero-age-main-sequence mass of 15--25\,$M_\odot$ that has lost a substantial fraction of its hydrogen envelope due to binary interaction \citep{Young_2006}. The distance to the SNR is estimated to be $\sim$\,3.4~kpc \citep{1995ApJ...440..706R}.
Cas~A exhibits a shell-like structure with an outer radius of $\sim$\,2.5~arcmin \citep[e.g.,][]{2003ApJ...589..818D}, dominated by nonthermal X-ray emission from accelerated particles \citep[e.g.,][]{2009ApJ...697..535P}.
On the other hand, the thermal X-rays from the SN ejecta are bright at the inner shell with an average radius of $\sim$\,1.6~arcmin \citep[e.g.,][]{2001ApJ...552L..39G}, where prominent emission lines of the intermediate-mass elements (IMEs: Si, S, Ar, Ca) and Fe have been detected \citep[e.g.,][]{1996A&A...307L..41V,2000ApJ...528L.109H,2012ApJ...746..130H}.
The morphology of the thermal X-ray emission is highly asymmetric and is particularly bright in the southeast (SE) and northwest (NW) regions. 
It is also known that the ejecta in the SE and NW regions are blueshifted and redshifted, respectively \citep[e.g.,][]{2001ApJ...560L.175H,2002A&A...381.1039W,2006ApJ...651..250L}.
A similar trend was confirmed in the optical and infrared wavelengths \citep[e.g.,][]{DeLaney_2010,Milisavljevic_2013,2024ApJ...965L..27M}.

Notably, Cas~A is known to host a central compact object or an X-ray faint NS, discovered by the Chandra X-ray Observatory \citep{Pavlov_2000}.
Measurement of its proper motion revealed that this NS is moving toward the south with a velocity of $\sim$\,430~km\,s$^{-1}$ \citep{2013arXiv1307.3539D,2024ApJ...962...82H}. 
It is suggested that the motion of the NS is physically related to the asymmetric SN explosion and resulting ejecta distribution. 
In fact, NuSTAR observations of radioactive $^{\rm 44}$Ti revealed that its spatial distribution is biased toward the north (i.e., opposite to the NS motion), with biased radial velocities of 1100--3000~km\,s$^{-1}$, i.e., only the redshift component exists \citep{Grefenstette_2014}. 
More recently, \cite{Katsuda_2018} investigated the spatial distribution of the IME ejecta and revealed that the center of their mass is shifted to the north with respect to the SN explosion center.
These results support a scenario that the NS kick is physically associated with asymmetric explosive mass ejection \citep{1994A&A...290..496J,1996PhRvL..76..352B,2017ApJ...842...13W}. 
However, measurement of the ejecta mass distribution is subject to uncertainty due to unshocked ejecta that cannot be observed in X-rays. 
In this work, we investigate the velocity distribution of the IME ejecta to compare it with the NS motion, using the X-ray microcalorimeter Resolve on board the X-ray Imaging and Spectroscopy Mission (XRISM: \citealp{10.1117/12.2565812}). 
With its unprecedented spectral resolution for extended sources, the Resolve has enabled us to measure the radial velocity of the plasma in an SNR with an accuracy of approximately 100~km\,s$^{-1}$ \citep[e.g.,][]{10.1093/pasj/psae080}, suitable for our objectives. 

This paper is organized as follows.
The observations and data reduction are described in Section~\ref{sec:2}. 
In Section~\ref{sec:3}, we present spectral analysis of the Resolve data and measure the radial velocity of the IME ejecta.
We then determine in Section~\ref{sec:4} the three-dimensional velocity of the ejecta with the aid of proper motion measurement using archival data of the Chandra X-ray Observatory. 
The results are discussed in Section~\ref{sec:5}.
Finally, we conclude this study in Section~\ref{sec:6}.
The errors quoted in the text and tables, and the error bars given in the figures represent the 1$\sigma$ confidence level, unless otherwise stated.

\section{Observations and Data Reduction}\label{sec:2}

XRISM observations of the SNR Cas~A were conducted twice in December 2023, the first on the southeast (SE) region (Observation ID: 000129000) and the second on the northwest (NW) region (Observation ID: 000130000), with the nominal aim points of (RA, Dec)$_{{\rm J}2000}$ = (350.83252, 58.79886) and (350.83062, 58.82244), respectively. 
The radial velocity component (toward Cas~A) of the Earth's orbital motion with respect to the Sun was about $-13$\,km\,s$^{-1}$ at the time of the observations.
The XRISM spacecraft is equipped with two instruments, Resolve \citep{10.1117/12.2630654} and Xtend \citep{10.1117/12.2626894,2025arXiv250208030N}, each of which consists of an X-ray Mirror Assembly (XMA) and a detector (an X-ray microcalorimeter array for the Resolve and X-ray CCDs for the Xtend) on the focal plane of the XMA.
The Resolve detector array consists of $6\times 6$ pixels, each with a size equivalent to a $0.\!'5 \times 0.\!'5$ square of the sky. 
Therefore, the field of view (FoV) of this instrument is $3' \times 3'$. 
During the observations, the aperture door with a $\sim$\,250\,$\mu$m thick beryllium window was closed, limiting the bandpass of the Resolve to energies above  $\sim$\,1.6\,keV. 
The Xtend consists of four CCD chips, making a wide FoV of $38' \times 38'$. This instrument was operating in the full-window mode during the observations.

\begin{figure}[t]
    \centering
    \includegraphics[width=0.96\linewidth]{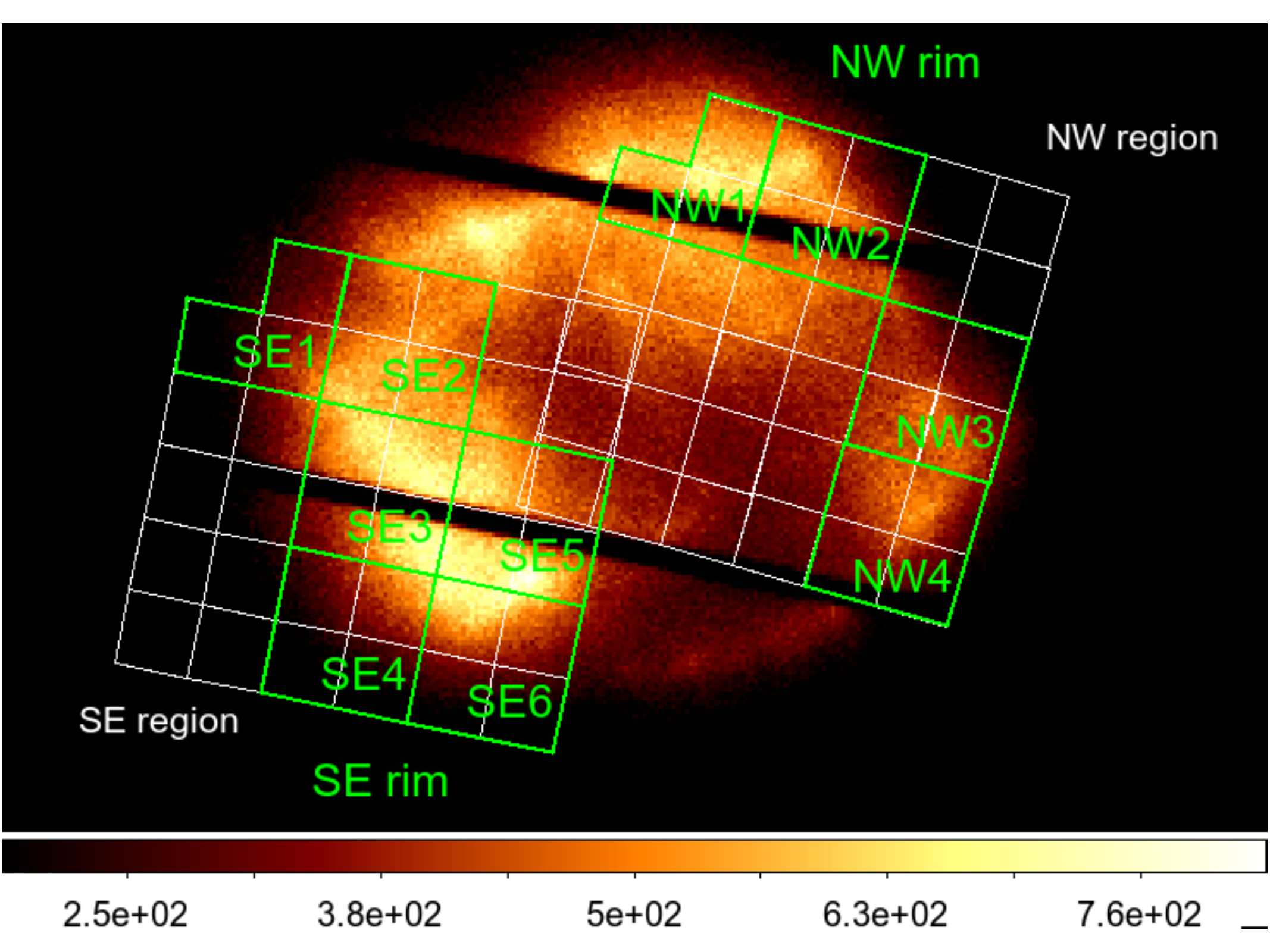}
    \vspace{0.3cm}
    \caption{XRISM/Xtend image of Cas~A in the 1.75--2.95~keV band. The white grids indicate the pixel map of the Resolve for the two observations. 
    The green boxes are where the Resolve spectra are extracted for detailed analysis.
    }
    \label{fig:an_re}
\end{figure}

Figure~\ref{fig:an_re} shows the Xtend image of Cas~A in 1.75--2.95\,keV, where the Resolve FoV and pixel positions of both observations are indicated by the white grids. 
Since our purpose is to accurately measure the radial velocity of the IME ejecta, we hereafter focus on analysis of the Resolve data, which have higher spectral resolution than the Xtend data. 
We reprocess the data using the HEASoft 6.34 software and the calibration database (CALDB) version 9, following the standard screening procedure \citep{mochizuki2025optimizationxrayeventscreening}. 
The resulting effective exposures are 182~ks and 166~ks for the SE and NW observations, respectively. 
We extract only Grade (ITYPE) 0 (High-resolution primary) events for spectral analysis.
Redistribution matrix files (RMFs) are generated using the \texttt{rslmkrmf} task with the L-size option, which means that the line-spread function contains the Gaussian core, exponential tail to low energy, escape peaks, and Si fluorescence. 
Auxiliary response files (ARFs) are generated using the \texttt{xaarfgen} task and a Chandra image in the 1.75--2.95~keV band extracted from ObsID = 4634, 4635, 4636, 4637, 4639, 5196 and 5319 data as an input sky image.

\section{Spectral Analysis}\label{sec:3}

We extract Resolve spectra from 10 regions indicated in Figure~\ref{fig:an_re} with the green boxes: 
SE1--SE6 from the SE observations and NW1--NW4 from the NW observation. 
Each of them contains 3 or 4 adjacent pixels, corresponding to a $1' \times 1'$ square region, the size comparable to the half-power diameter (HPD) of the XMA's point spread function (PSF).

In the following subsections, we first fit the 1.75--2.95~keV spectra (containing the K-shell emission of Si and S) with an ad hoc, single-temperature plasma model, revealing the presence of complex velocity structure that cannot be explained by such a simple model even in this narrow energy band (\S3.1). 
We then measure the centroid energies of the emission lines using Gaussian models to determine their offset from the theoretical rest-frame energies, obtaining different velocity shift values between the He$\alpha$ and Ly$\alpha$ emissions (\S3.2). 
Lastly, we introduce a model consisting of two plasmas with different radial velocities, electron temperatures, and ionization parameters that reasonably reproduces the observed spectra in the 1.75--2.95\,keV band (\S3.3).
The optimal binning method \citep{2016A&A...587A.151K} is applied to all the spectra using the \texttt{ftgrouppha} task in FTOOLS. We ignore non X-ray background (NXB), since its contribution to the observed spectra is negligible in the energy band we analyze. 
The spectral fitting is performed based on the $C$-statistic \citep{1979ApJ...228..939C}, using the XSPEC software version 12.14.1 \citep{1996ASPC..101...17A}.

\subsection{Single component modeling}\label{sec:3.1}

We start the spectral fitting with a single component of a \texttt{bvvrnei} model, which reproduces thermal emission from an optically-thin plasma in non-equilibrium ionization (NEI). Note that previous study with XMM-Newton indicated that the spectra of Cas~A in the Si and S K band were well modeled with a single NEI component \citep[e.g.,][]{2002A&A...381.1039W}. 
The free parameters are the electron temperature ($kT_\mathrm{e}$), Si and S abundances relative to the solar values of \citet{2000ApJ...542..914W}, ionization timescale ($\tau=n_et$), redshift ($z$), velocity dispersion ($\sigma_v$), and normalization. 
The abundances of P and Cl are tied to the Si abundance, whereas those of the other elements are fixed to the solar values of \cite{2000ApJ...542..914W}. 
The initial (pre-shock) temperature is fixed to 0.01\,keV, since the plasma in Cas~A is known to be ionizing. 
The hydrogen column density of foreground absorption (\texttt{tbabs}) is fixed to $\mathrm{1.3\times10^{22}}$~cm$^{-2}$ (\citealp{2012ApJ...746..130H}; \citealp{2024ApJ...969..155W}). 
This model yields the best-fit values of electron temperature and redshift (or blueshift when the value is negative) given in Table~\ref{tab:1nei_redshift} and the other parameters given in Table~\ref{tab:1nei} in Appendix. The spectra of all the regions fitted with this single component model are also shown in Appendix (Figure~\ref{fig_1nei_AL}). 
We find that the IME ejecta have negative radial velocity (blueshifted) in the SE regions, whereas those in the NW regions have positive values (redshifted), confirming the previous studies with traditional X-ray CCDs \citep[e.g.,][]{2001ApJ...560L.175H,2002A&A...381.1039W}. 

\begin{figure}[t]
    \centering
    \includegraphics[width=0.96\linewidth]{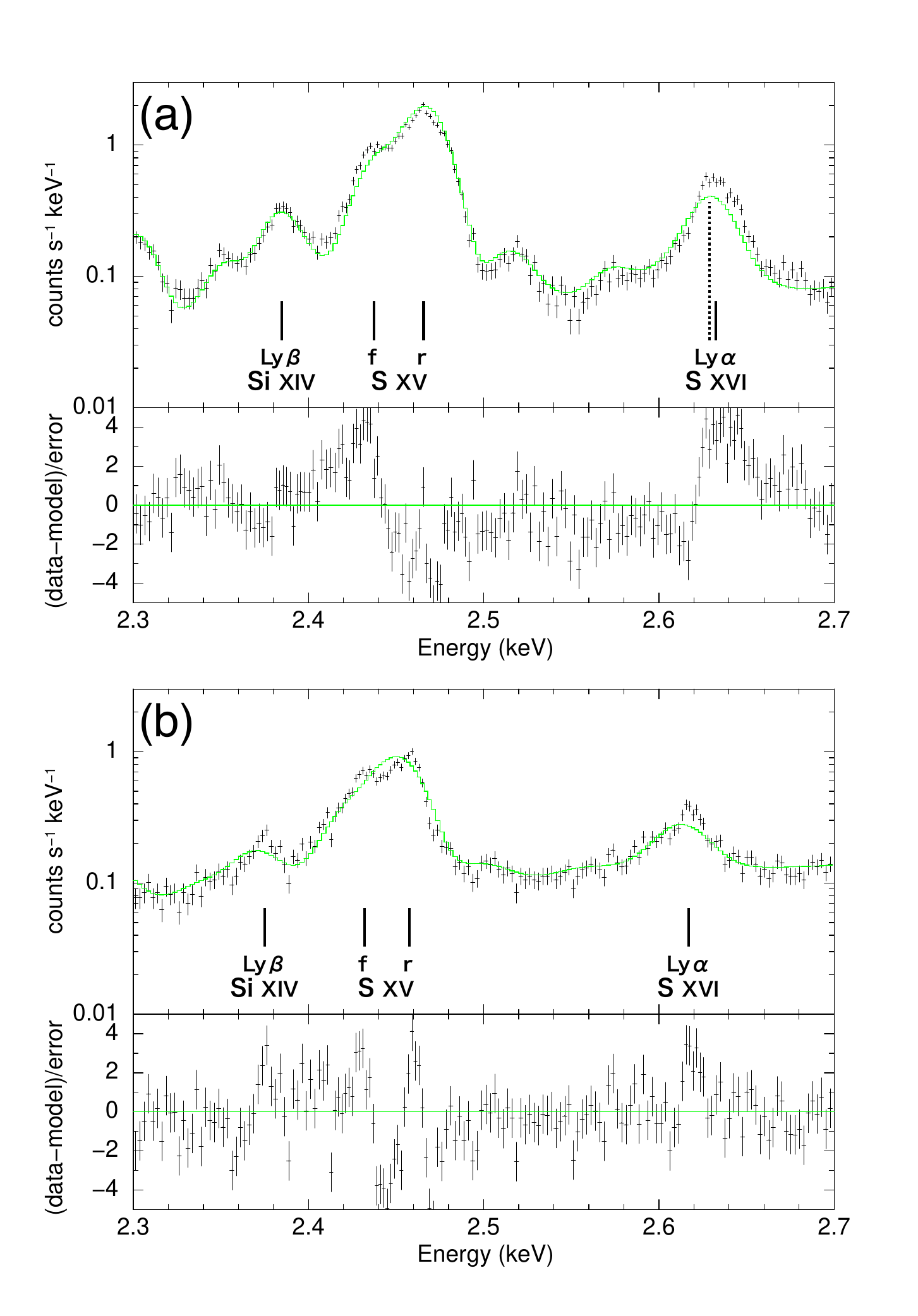}
    \caption{(a) Resolve spectrum in the 2.3--2.7~keV band of Region SE3 fitted with a single NEI model (green). The dashed line indicates the central energy of the S~\emissiontype{XVI} Ly$\alpha$ line derived from the best-fit model. \ (b) Same as Panel (a) but of Region NW3.
    }
    \label{fig_1nei}
\end{figure}

Figure~\ref{fig_1nei}a and \ref{fig_1nei}b show the 2.3--2.7\,keV spectra of Region SE3 and NW3, respectively, as representatives of typical fitting results. 
In the SE3 spectrum, positive residuals are prominent at the low-energy side of the S~\emissiontype{XV} (He$\alpha$) emission and the high-energy side of the S~\emissiontype{XVI} (Ly$\alpha$) emission.
This implies that the energy shift from the rest-frame value is overestimated for the He$\alpha$ emission but underestimated for the Ly$\alpha$ emission by this single-component model. We obtain similar results from the other SE regions as well as the NW1 and NW2 regions (see Appendix). 
In the NW3 spectrum (Figure~\ref{fig_1nei}b), on the other hand, the observed energy shift is relatively well modeled for both He$\alpha$ and Ly$\alpha$ lines. However, the line profiles are not perfectly reproduced; the observed profiles seem to be composed of a `narrow' component, in addition to the broadened emission that is reproduced by the single-component model. 
A similar result is also obtained from the NW4 region. 

\begin{table}
     \tbl{Electron temperature and redshift obtained with a single NEI model.}{%
      \begin{tabular}{lcc}
      \hline
      \hline
       & $kT_\mathrm{e}$           & $v_{\rm shift}$         \\
Region & (keV)                     & (km~s$^{-1}$)       \\ \hline
SE1    & $1.37^{+0.05}_{-0.02}$    & $-953^{+30}_{-35}$  \\
SE2    & $1.21^{+0.02}_{-0.03}$    & $-697\pm18$         \\
SE3    & $1.373^{+0.005}_{-0.002}$ & $-913\pm12$         \\
SE4    & $1.40\pm0.04$             & $-906\pm19$         \\
SE5    & $1.31^{+0.03}_{-0.02}$    & $-1188^{+15}_{-16}$ \\
SE6    & $1.41^{+0.04}_{-0.03}$    & $-1026^{+18}_{-19}$ \\
NW1    & $1.64^{+0.05}_{-0.02}$    & $1444^{+27}_{-20}$  \\
NW2    & $1.64^{+0.02}_{-0.05}$    & $1504^{+19}_{-20}$  \\
NW3    & $1.71^{+0.03}_{-0.06}$    & $1028^{+28}_{-33}$  \\
NW4    & $1.70^{+0.04}_{-0.08}$    & $1005^{+31}_{-36}$ \\ \hline
      \end{tabular}
      }
      \label{tab:1nei_redshift}
\end{table}

\subsection{Gaussian modeling to the emission lines}\label{sec:3.2}

\begin{table*}
  \tbl{Best-fit parameters of strong lines in the 1.75--2.1~keV and 2.35--2.8~keV bands.}{%
      \begin{tabular}{lccccccccc}
      \hline
\hline
       &                    & width$^{\dag}$                & $v_{\rm shift}$                  & \multicolumn{2}{c}{normalization ($10^{-2}$)}   &                     & width$^{\dag}$                & $v_{\rm shift}$                  &                           \\
Region & line               & (km\,s$^{-1}$)                 & (km\,s$^{-1}$)          & $f^*$                 & $r^*$                  & line                & (km\,s$^{-1}$)                 & (km\,s$^{-1}$)          & normalization$^{**}$ ($10^{-2}$) \\ \hline
\multirow{2}{*}{SE1}    & Si~\emissiontype{XIII}~He$\alpha$ & $1535^{+89}_{-78}$  & $-790^{+124}_{-117}$ & $5.9\pm0.6$    & $12.6\pm0.7$   & Si~\emissiontype{XIV}~Ly$\alpha$ & $1476^{+70}_{-67}$  & $-1192^{+75}_{-74}$ & $4.8\pm0.2$       \\
       & S~\emissiontype{XV}~He$\alpha$   & $1512^{+52}_{-49}$ & $-895^{+84}_{-80}$ & $2.1\pm0.1$    & $3.4\pm0.1$    & S~\emissiontype{XVI}~Ly$\alpha$   & $1582^{+82}_{-78}$ & $-1218^{+87}_{-86}$ & $0.78\pm0.04$       \\ 
\multirow{2}{*}{SE2}    & Si~\emissiontype{XIII}~He$\alpha$ & $1545^{+42}_{-40}$  & $-706^{+71}_{-69}$ & $8.4\pm0.4$    & $14.5\pm+0.5$   & Si~\emissiontype{XIV}~Ly$\alpha$ & $1469^{+50}_{-49}$  & $-1029^{+56}_{-55}$ & $3.0\pm0.1$       \\
       & S~\emissiontype{XV}~He$\alpha$   & $1615^{+32}_{-31}$ & $-492^{+54}_{-52}$ & $2.04\pm0.07$ & $3.38\pm0.08$ & S~\emissiontype{XVI}~Ly$\alpha$   & $1592^{+69}_{-66}$ & $-999^{+74}_{-73}$ & $0.45\pm0.02$    \\
\multirow{2}{*}{SE3}    & Si~\emissiontype{XIII}~He$\alpha$ & $1467^{+39}_{-37}$  & $-717^{+64}_{-63}$ & $5.5\pm0.3$    & $14.0\pm0.6$   & Si~\emissiontype{XIV}~Ly$\alpha$ & $1290\pm32$  & $-1150^{+36}_{-35}$ & $4.0\pm{+0.1}$       \\
       & S~\emissiontype{XV}~He$\alpha$   & $1348^{+18}_{-17}$ & $-691^{+25}_{-24}$ & $1.84\pm0.04$ & $3.69\pm0.05$ & S~\emissiontype{XVI}~Ly$\alpha$   & $1248^{+37}_{-36}$ & $-1225^{+38}_{-37}$ & $0.62\pm0.02$      \\
\multirow{2}{*}{SE4}    & Si~\emissiontype{XIII}~He$\alpha$ & $1345^{+59}_{-54}$  & $-724^{+72}_{-70}$ & $5.2\pm0.4$    & $12.1\pm0.5$   & Si~\emissiontype{XIV}~Ly$\alpha$ & $1237^{+52}_{-50}$  & $-1056\pm55$ & $4.1\pm0.2$       \\
       & S~\emissiontype{XV}~He$\alpha$   &$1211^{+26}_{-25}$ & $-723^{+35}_{-34}$ & $1.48\pm0.06$ & $3.27\pm0.07$ & S~\emissiontype{XVI}~Ly$\alpha$   & $1258^{+63}_{-61}$ & $-1024^{+64}_{-63}$ & $0.61\pm0.03$    \\
\multirow{2}{*}{SE5}    & Si~\emissiontype{XIII}~He$\alpha$ & $1435^{+35}_{-33}$  & $-1198^{+52}_{-51}$ & $6.7\pm0.3$    & $14.9\pm0.4$   & Si~\emissiontype{XIV}~Ly$\alpha$ & $1485^{+52}_{-50}$ & $-1254^{+54}_{-53}$ & $2.9\pm0.1$       \\
       & S~\emissiontype{XV}~He$\alpha$   & $1586^{+29}_{-28}$ & $-942^{+49}_{-47}$ & $1.87\pm0.07$ & $3.61\pm0.08$ & S~\emissiontype{XVI}~Ly$\alpha$   & $1529^{+79}_{-75}$ & $-1257^{+79}_{-78}$ & $0.37\pm0.02$    \\ 
\multirow{2}{*}{SE6}    & Si~\emissiontype{XIII}~He$\alpha$ & $1358^{+45}_{-39}$  & $-976^{+65}_{-62}$ & $6.7^{+0.5}_{-0.4}$    & $16.0^{+0.6}_{-0.5}$   & Si~\emissiontype{XIV}~Ly$\alpha$ & $1415\pm60$  & $-1132^{+68}_{-65}$ & $3.5^{+0.1}_{-0.2}$       \\
       & S~\emissiontype{XV}~He$\alpha$   & $1393^{+29}_{-28}$ & $-828^{+43}_{-42}$ & $1.85\pm0.07$ & $3.73\pm0.08$ & S~\emissiontype{XVI}~Ly$\alpha$   & $1360^{+84}_{-79}$ & $-942\pm78$ & $0.47^{+0.03}_{-0.02}$    \\
\multirow{2}{*}{NW1}    & Si~\emissiontype{XIII}~He$\alpha$ & $1929^{+71}_{-65}$ & $1085^{+119}_{-114}$  & $7.4\pm0.7$    & $18.6\pm0.8$   & Si~\emissiontype{XIV}~Ly$\alpha$ & $2030^{+67}_{-64}$ & $495^{+78}_{-77}$  & $3.6\pm0.1$       \\
       & S~\emissiontype{XV}~He$\alpha$   & $2311\pm50$ & $2093^{+77}_{-82}$  & $1.3\pm0.1$    & $5.5^{+0.5}_{-0.2}$    & S~\emissiontype{XVI}~Ly$\alpha$   & $1832^{+78}_{-74}$ & $1838\pm76$  & $0.58\pm0.02$    \\
\multirow{2}{*}{NW2}    & Si~\emissiontype{XIII}~He$\alpha$ & $1784^{+82}_{-73}$ & $1190^{+129}_{-116}$  & $7.1\pm0.7$    & $17.6^{+0.9}_{-0.8}$   & Si~\emissiontype{XIV}~Ly$\alpha$ & $1615^{+59}_{-57}$ & $1333\pm64$  & $3.7\pm0.1$       \\
       & S~\emissiontype{XV}~He$\alpha$   & $1825^{+41}_{-39}$ & $1887^{+57}_{-56}$  & $1.9\pm0.1$    & $5.4\pm0.1$    & S~\emissiontype{XVI}~Ly$\alpha$   & $1721^{+68}_{-65}$ & $1914^{+66}_{-65}$  & $0.76\pm0.03$    \\ 
\multirow{2}{*}{NW3}    & Si~\emissiontype{XIII}~He$\alpha$ & $1240^{+86}_{-36}$  & $476^{+91}_{-83}$  & $5.2\pm0.4$    & $6.2^{+0.5}_{-0.4}$    & Si~\emissiontype{XIV}~Ly$\alpha$ & $1451^{+76}_{-72}$  & $844^{+78}_{-77}$  & $2.7\pm0.1$       \\
       & S~\emissiontype{XV}~He$\alpha$   & $1093^{+40}_{-39}$  & $480\pm51$  & $1.20\pm0.05$ & $1.79\pm0.08$ & S~\emissiontype{XVI}~Ly$\alpha$   & $1457^{+100}_{-99}$ & $730\pm89$  & $0.74\pm0.04$    \\ 
\multirow{2}{*}{NW4}    & Si~\emissiontype{XIII}~He$\alpha$ & $1365^{+70}_{-65}$  & $523^{+80}_{-77}$  & $5.5\pm0.3$    & $7.8\pm0.4$    & Si~\emissiontype{XIV}~Ly$\alpha$ & $1551^{+77}_{-75}$ & $892^{+84}_{-83}$  & $2.8\pm0.1$       \\ 
       & S~\emissiontype{XV}~He$\alpha$   & $1642^{+116}_{-130}$ & $1047^{+190}_{-189}$  & $1.5\pm0.1$    & $3.0\pm0.4$    & S~\emissiontype{XVI}~Ly$\alpha$   & $1357^{+107}_{-111}$ & $838^{+89}_{-91}$  & $0.48\pm0.03$    \\ \hline
\end{tabular}
}
  \begin{tabnote}
    \footnotemark[$*$]  $f$ and $r$ represent the forbidden and resonance lines of the He$\alpha$ emission, respectively.\\
    \footnotemark[$**$] The values are given for the sum of the Ly$\alpha_1$ and Ly$\alpha_2$ fluxes.\\
    \footnotemark[$\dag$] The values are 1$\sigma$ velocity dispersion.
  \end{tabnote}
  \label{tab:zgauss_all}
\end{table*}

\begin{figure}[t]
    \centering
    \includegraphics[width=0.96\linewidth]{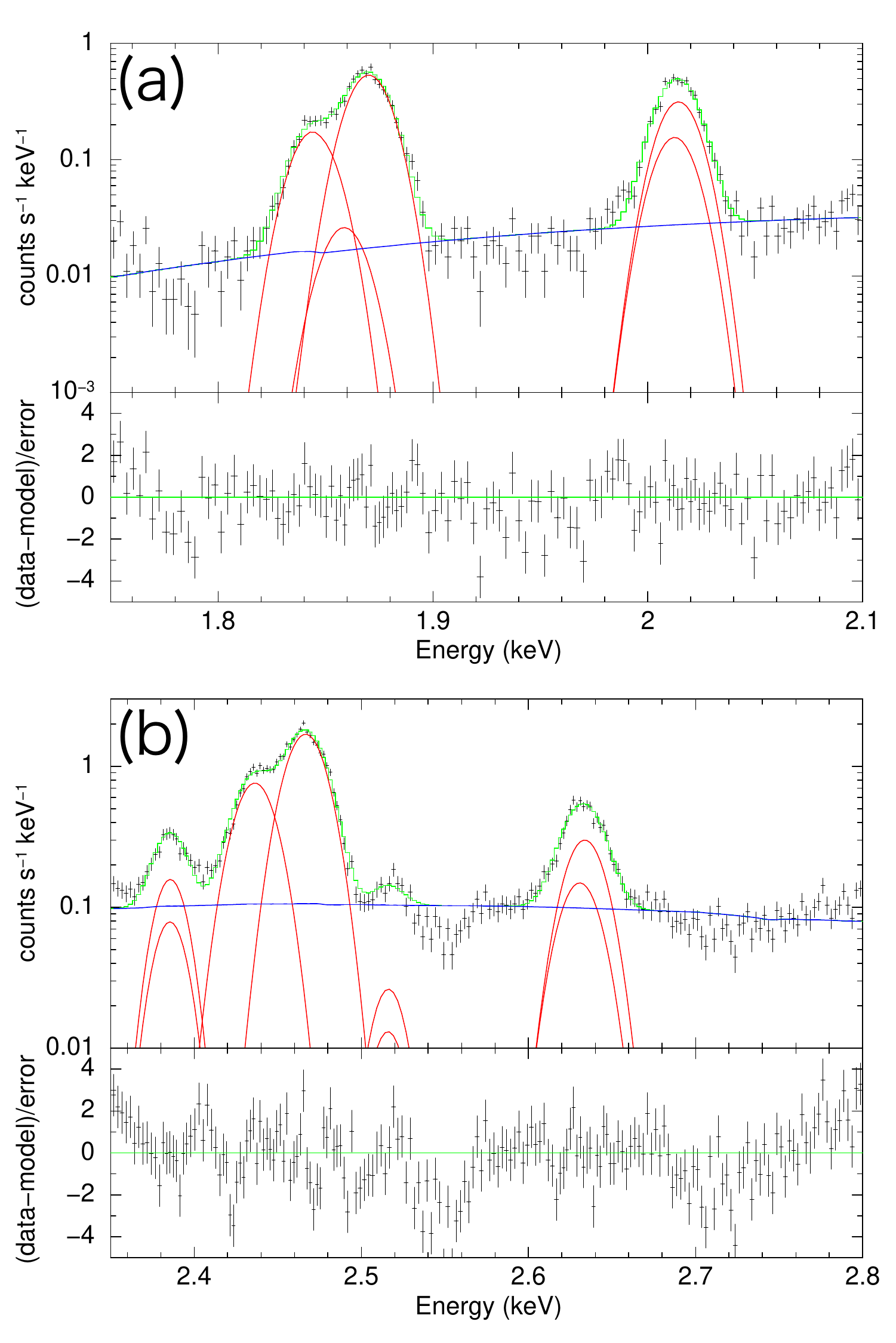}
    \caption{(a) Resolve spectrum in the 1.75--2.1~keV band from Region SE3 fitted with the Gaussian functions (red) and bremsstrahlung continuum (blue). \ 
    (b) Same as Panel (a) but in the 2.35--2.8~keV band.
    }
    \label{fig:zgauss_fig}
\end{figure}

\begin{figure}[t]
    \centering
    \includegraphics[width=0.96\linewidth]{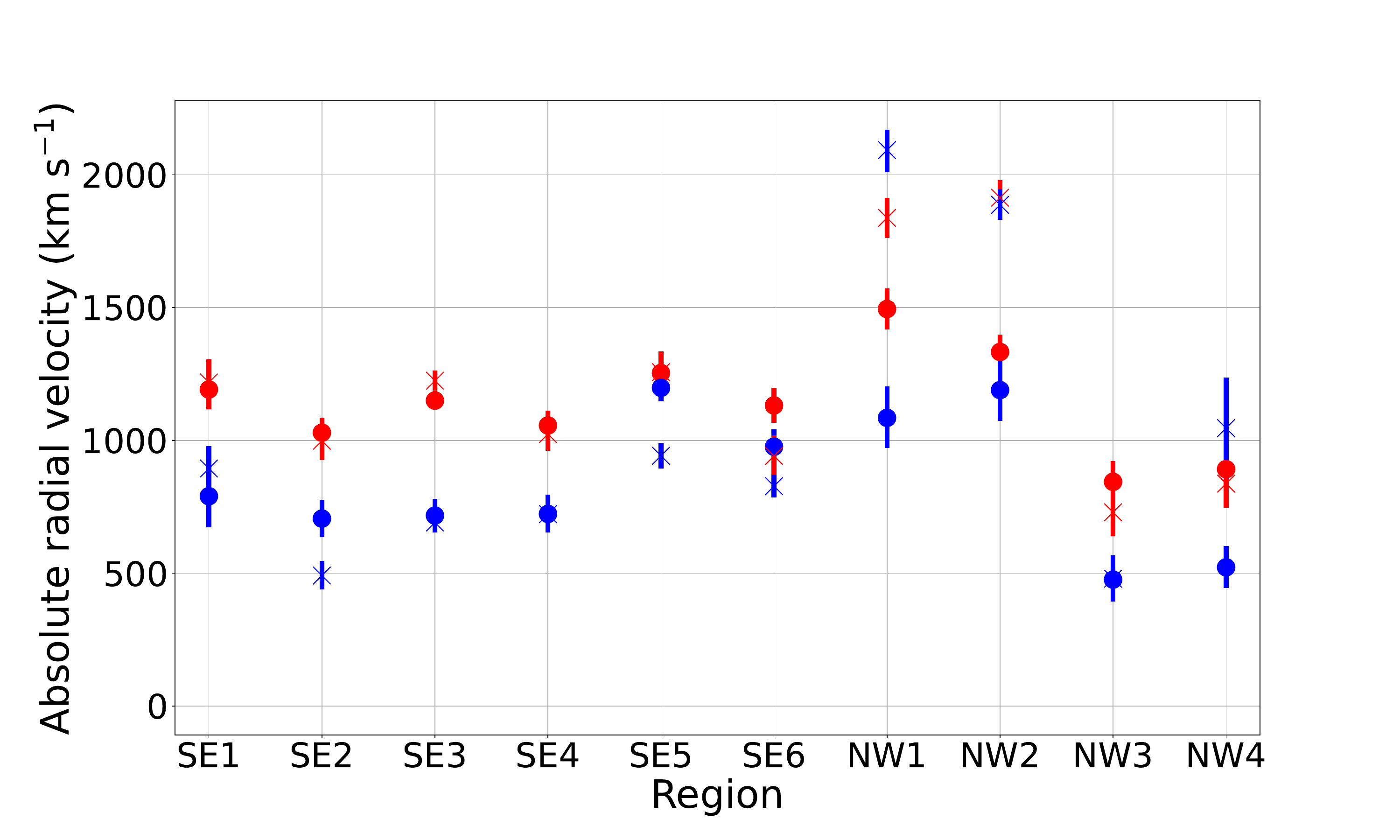}
    \caption{Doppler velocities of Si~\emissiontype{XIV} Ly$\alpha$ (red dot), Si~\emissiontype{XIII} He$\alpha$ (blue dot), S~\emissiontype{XVI}~Ly$\alpha$ (red cross) and S~\emissiontype{XV} He$\alpha$ (blue cross), derived by the Gaussian modeling in Section 3.2.
    }
    \label{fig:zgauss_dop}
\end{figure}

Our ad hoc analysis in the previous subsection has indicated that more than one plasma component with different radial velocities contribute to the spectrum of each region.
This result motivates us to fit each emission line with a Gaussian function to determine the redshift or blueshift values of He$\alpha$ and Ly$\alpha$ lines independently. 
We first analyze the spectra in the 1.75--2.1\,keV band (Figure~\ref{fig:zgauss_fig}a for SE3 regions as a representative), where the forbidden ($f$), intercombination ($i$), and resonance ($r$) lines of Si~\emissiontype{XIII}, as well as the Ly$\alpha_1$ and Ly$\alpha_2$ lines of Si~\emissiontype{XIV} are detected. 
Each of these lines is fitted with a \texttt{zgauss} model, with the centroid energy fixed to their theoretical rest frame value (referring to AtomDB). 
The line width and redshift (or blueshift) are treated as free parameters, but those of the He$\alpha$ lines (i.e., $f$, $i$, and $r$) are linked to one another, and those of the Ly$\alpha_1$ and Ly$\alpha_2$ lines are linked to each other. 
We also fix the Ly$\alpha_1$/Ly$\alpha_2$ ratio to 2, i.e., statistical weight ratio between the excited states. 
The continuum is fitted with a \texttt{bremsstrahlung} model. 
We subsequently analyze the 2.35--2.8\,keV spectra (i.e., S K band; Figure~\ref{fig:zgauss_fig}b for SE3 regions) with a similar approach. 
This energy band contains not only the S~\emissiontype{XV}~He$\alpha$ and \emissiontype{XVI}~Ly$\alpha$ emissions but also the Si~\emissiontype{XIV} Ly$\beta$ and Ly$\gamma$ emissions. 
Since their photon statistics are relatively low, we fix the width and redshift of the Ly$\beta$ and Ly$\gamma$ lines to the values obtained for the Si~\emissiontype{XIV}~Ly$\alpha$ emission. The ratios of Ly$\beta_1$/Ly$\beta_2$ and Ly$\gamma_1$/Ly$\gamma_2$ are also fixed to 2.

\begin{table*}
  \tbl{Best-fit parameters of the two component NEI models.}{%
    \begin{tabular}{lccccccccc}
      \hline
      \hline
       & & $kT_\mathrm{e}$& Si abundance& S abundance& $\tau$& $v_{\rm shift}$& $\sigma_v$& & \\
      Region & component & (keV)& (solar)& (solar)& (10$^{11}$~cm$^{-3}$~s)& (km\,s$^{-1}$)& (km\,s$^{-1}$) & normalization & $C$-stat/dof\\ 
      \hline
\multirow{2}{*}{SE1}& low-$\tau$&$5.8^{+8.6}_{-2.2}$& \multirow{2}{*}{$5.7^{+0.3}_{-0.2}$}& \multirow{2}{*}{$4.2\pm0.2$}&$0.74^{+0.25}_{-0.17}$&$-2025^{+41}_{-43}$&$638^{+66}_{-79}$&$0.18^{+0.04}_{-0.03}$&\multirow{2}{*}{$453.56/338$} \\
& high-$\tau$&$1.23^{+0.09}_{-0.05}$& & &$1.91^{+0.29}_{-0.28}$&$-330\pm85$&$1724^{+48}_{-47}$&$1.47^{+0.12}_{-0.14}$& \\
\multirow{2}{*}{SE2}& low-$\tau$&$1.36^{+0.03}_{-0.08}$& \multirow{2}{*}{$5.5^{+0.2}_{-0.1}$}& \multirow{2}{*}{$4.5^{+0.2}_{-0.1}$}&$0.21\pm0.02$&$-451^{+69}_{-90}$&$1586^{+41}_{-40}$&$0.97^{+0.10}_{-0.07}$&\multirow{2}{*}{$836.51/375$} \\
& high-$\tau$&$1.8\pm0.1$& & &$1.40^{+0.12}_{-0.18}$&$-1187^{+50}_{-39}$&$1303^{+31}_{-27}$&$0.62^{+0.04}_{-0.05}$& \\
\multirow{2}{*}{SE3}& low-$\tau$&$1.18^{+0.07}_{-0.06}$& \multirow{2}{*}{$6.4^{+0.2}_{-0.1}$}& \multirow{2}{*}{$5.1\pm0.1$}&$0.32\pm0.02$&$-445^{+83}_{-66}$&$1180^{+63}_{-75}$&$0.76^{+0.07}_{-0.08}$&\multirow{2}{*}{$742.83/474$} \\
& high-$\tau$&$2.1\pm0.1$& & &$1.29^{+0.10}_{-0.09}$&$-1205^{+26}_{-29}$&$1260^{+17}_{-14}$&$0.60\pm0.05$& \\
\multirow{2}{*}{SE4}& low-$\tau$&$1.0\pm0.1$& \multirow{2}{*}{$4.9\pm0.2$}& \multirow{2}{*}{$3.8\pm0.1$}&$0.23\pm0.03$&$-188^{+86}_{-90}$&$782^{+80}_{-70}$&$0.71^{+0.26}_{-0.15}$&\multirow{2}{*}{$474.02/347$} \\
& high-$\tau$&$1.8^{+0.12}_{-0.05}$& & &$1.36\pm0.13$&$-1059^{+28}_{-29}$&$1186^{+22}_{-20}$&$1.02^{+0.06}_{-0.08}$& \\
\multirow{2}{*}{SE5}& low-$\tau$&$1.47^{+0.10}_{-0.09}$& \multirow{2}{*}{$4.5\pm0.1$}& \multirow{2}{*}{$3.56^{+0.09}_{-0.08}$}&$0.20^{+0.02}_{-0.01}$&$-1250^{+66}_{-74}$&$1430^{+51}_{-46}$&$0.96^{+0.10}_{-0.09}$&\multirow{2}{*}{$863.40/480$} \\
& high-$\tau$&$1.9^{+0.2}_{-0.1}$& & &$1.02^{+0.09}_{-0.13}$&$-1423^{+36}_{-33}$&$1374^{+26}_{-14}$&$0.79^{+0.06}_{-0.08}$& \\
\multirow{2}{*}{SE6}& low-$\tau$&$1.3\pm0.1$& \multirow{2}{*}{$4.6\pm0.2$}& \multirow{2}{*}{$3.6\pm0.1$}&$0.22^{+0.03}_{-0.02}$&$-1117^{+70}_{-69}$&$1231^{+52}_{-47}$&$1.02^{+0.16}_{-0.15}$&\multirow{2}{*}{$554.92/416$} \\
& high-$\tau$&$2.4\pm0.2$& & &$0.82^{+0.08}_{-0.07}$&$-1112^{+30}_{-31}$&$1288\pm26$&$0.76\pm0.07$& \\
\multirow{2}{*}{NW1}& low-$\tau$&$1.14^{+0.10}_{-0.09}$& \multirow{2}{*}{$6.3^{+0.2}_{-0.1}$}& \multirow{2}{*}{$5.3\pm0.2$}&$0.22\pm0.02$&$449^{+113}_{-109}$&$1905^{+64}_{-63}$&$1.16^{+0.14}_{-0.06}$&\multirow{2}{*}{$598.19/378$} \\
& high-$\tau$&$3.4^{+0.5}_{-0.3}$& & &$0.64^{+0.04}_{-0.05}$&$1652^{+44}_{-42}$&$1994^{+35}_{-37}$&$0.48\pm0.04$& \\
\multirow{2}{*}{NW2}& low-$\tau$&$1.19^{+0.10}_{-0.07}$& \multirow{2}{*}{$4.2\pm0.1$}& \multirow{2}{*}{$3.8\pm0.1$}&$0.26^{+0.02}_{-0.03}$&$889^{+76}_{-89}$&$1717^{+52}_{-57}$&$1.65^{+0.14}_{-0.18}$&\multirow{2}{*}{$549.54/381$} \\
& high-$\tau$&$3.3^{+0.7}_{-0.3}$& & &$0.77^{+0.07}_{-0.09}$&$1599^{+42}_{-36}$&$1614^{+35}_{-33}$&$0.71^{+0.07}_{-0.09}$& \\
      \hline
    \end{tabular}
  }
  \label{tab:2nei_res}
\end{table*}

\begin{figure}[t]
    \centering
    \includegraphics[width=0.96\linewidth]{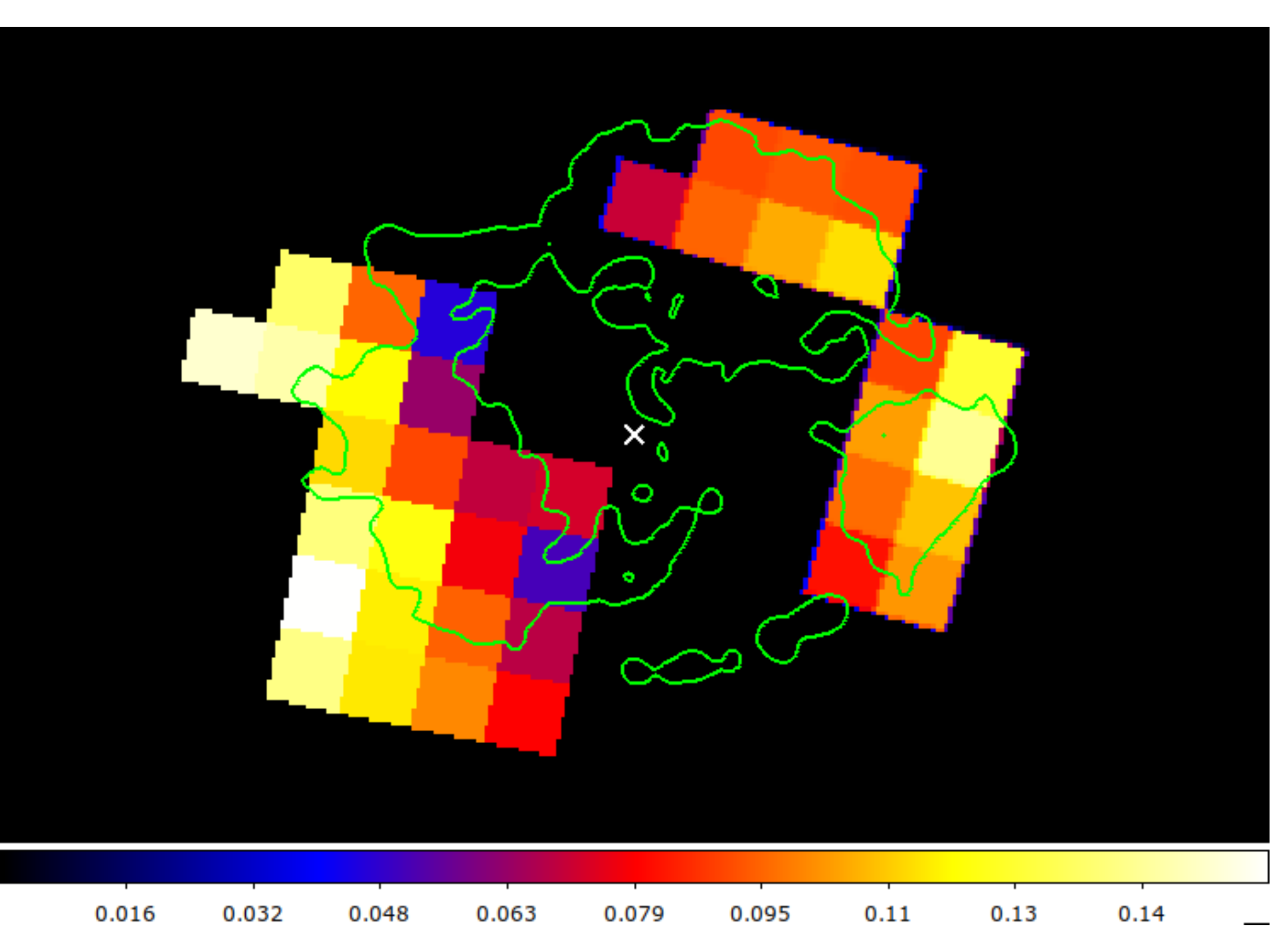}
    \caption{The flux ratio map of S~\emissiontype{XVI}~Ly$\alpha$ / S~\emissiontype{XV} He$\alpha$. The contours show the Chandra 1.75--2.95~keV image. The white cross shows the center of expansion, at (R.A., Dec.) = (23h23m27.77s, 58d48m49.4s) (\citealp{2001AJ....122..297T}).
    }
    \label{fig:color_map}
\end{figure}

The best-fit parameters and models from this analysis are given in Table~\ref{tab:zgauss_all} and Figure~\ref{fig:zgauss_all} in Appendix, respectively. 
Figure~\ref{fig:zgauss_dop} shows the absolute values of the radial velocities for the ten regions, converted from the best-fit redshift values. 
We confirm that the velocity derived from the Ly$\alpha$ emission is higher than those from the He$\alpha$ emission in most regions. 
It would also be worth noting that the measured line width is systematically broader in the NW regions than in the SE regions (Table~\ref{tab:zgauss_all}), suggesting that the velocity dispersion of the IME ejecta is larger in the former, consistent with the optical/infrared observations of the ejecta knot \citep{Milisavljevic_2013,DeLaney_2010}.
More detailed analysis of the velocity dispersion observed in the Resolve data is presented in a separate paper (Vink et al.\ 2025).
We also show in Figure~\ref{fig:color_map} the ratio between the S~\emissiontype{XVI}~Ly$\alpha$ and S~\emissiontype{XV} He$\alpha$ flux (i.e., Gaussian normalization) in each pixel, the interpretation of which will be discussed later. 

\subsection{Two-temperature plasma model}\label{sec:3.3}

\begin{figure*}
    \centering
    \includegraphics[width=0.85\linewidth]{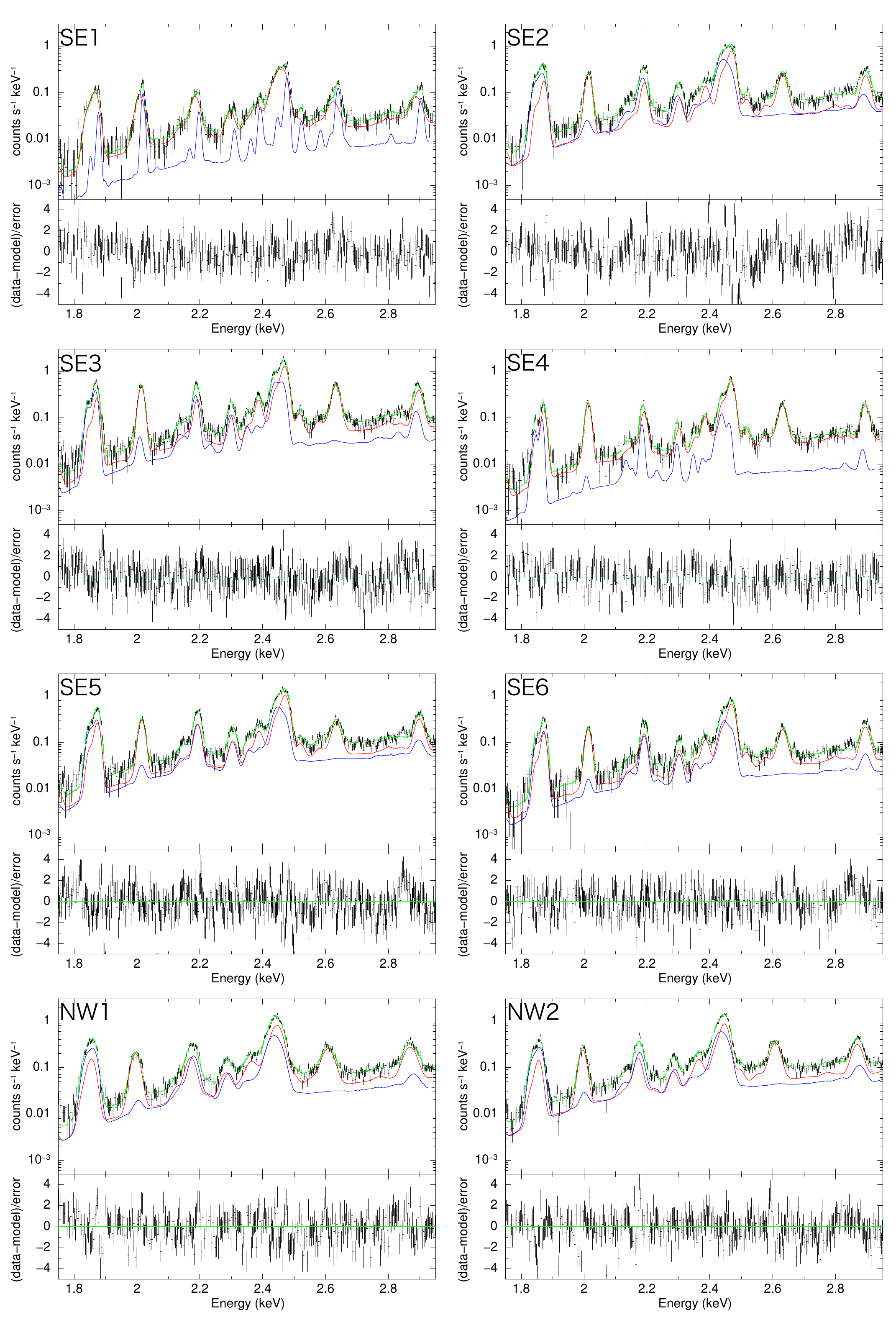}
    \caption{Resolve spectra in the 1.75--2.95~keV band from Regions SE1--SE6, NW1, NW2, fitted with the two-component NEI model. 
    The low-$\tau$ and high-$\tau$ components of the best-fit model are indicated as blue and red, respectively.
    }
    \label{fig:2nei_all}
\end{figure*}

As the final step of our spectral analysis, we introduce a more physically realistic model, taking into account the results of the previous subsections: presence of multiple plasma components with different radial velocity and ionization timescale, inferred from the line centroids of the He$\alpha$ and Ly$\alpha$ emissions.
We fit the spectra with two \texttt{bvvrnei} components with foreground absorption (\texttt{tbabs} with $N_{\rm H} = \mathrm{1.3\times10^{22}}$~cm$^{-2}$). 
The electron temperature ($kT_\mathrm{e}$), ionization timescale ($\tau=n_et$), redshift ($z$), velocity dispersion ($\sigma_v$), and normalization are allowed to vary independently between the two components. 
However, the abundances of Si and S are tied between the two components, since our model is unable to constrain them independently (i.e., if the abundances of each component are fitted independently, the constrained error ranges of several parameters become extremely large). 
Furthermore, the abundances of P and Cl are linked to the abundance of Si in both NEI components. 


\begin{figure}[t]
   \centering
   \includegraphics[width=0.96\linewidth]{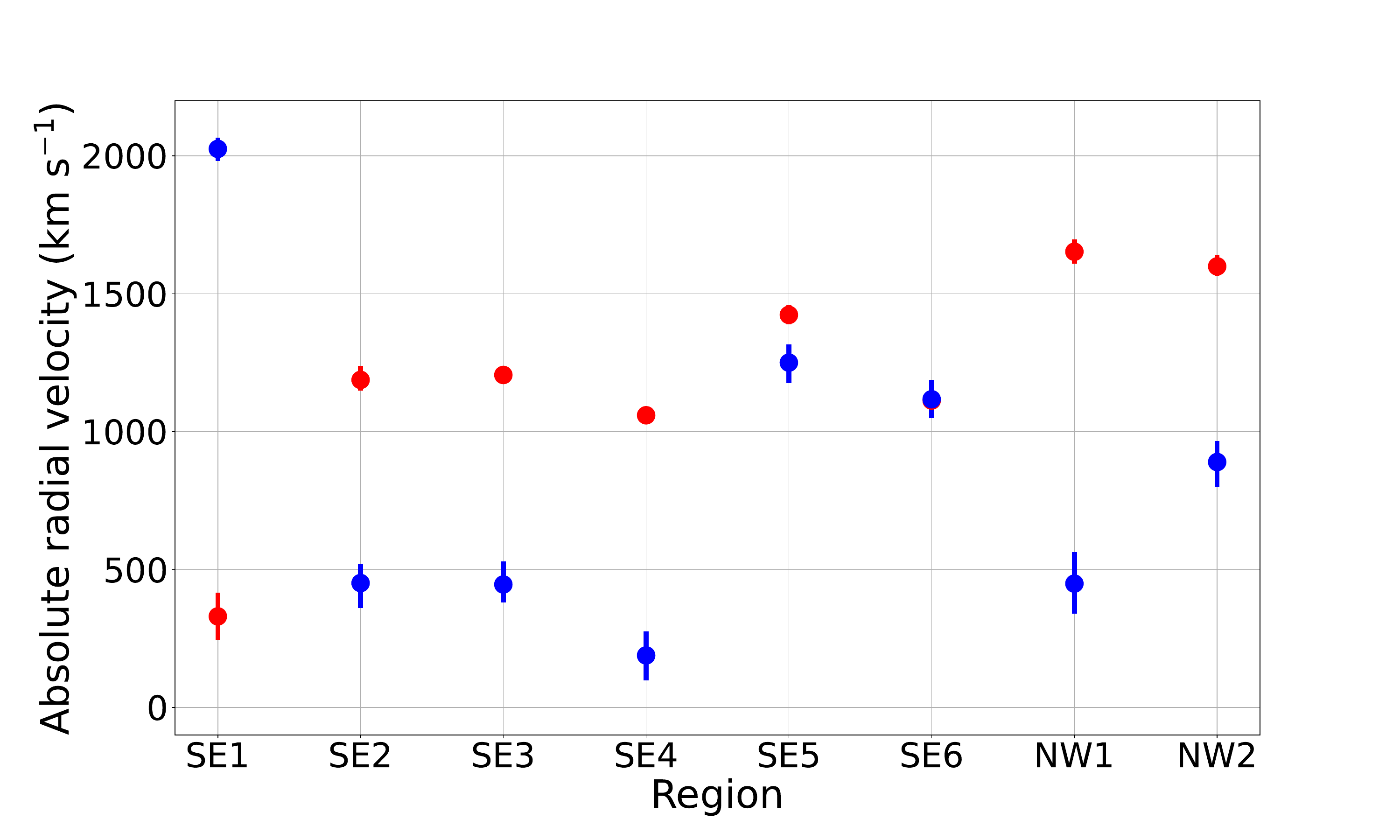}
   \caption{Absolute values of the radial velocities of the low-$\tau$ and high-$\tau$ component (blue and red, respectively) in each region.
   }
   \label{fig:doppler_2nei}
\end{figure}

This model yields the best-fit results given in Table~\ref{tab:2nei_res} and Figure~\ref{fig:2nei_all}, where the NW3 and NW4 regions are excluded, because the presence of narrow emission components in their spectra (\S3.1) requires even more complicated models including possible contributions of charge exchange emission. We will report detailed analysis of these regions in a separate paper (Sonoda et al.\ in preparation).
For the other eight regions, significantly different values of $\tau$ and $z$ are obtained between the two components, as expected.
Figure~\ref{fig:doppler_2nei} shows the absolute values of the radial velocities for both components in each region, confirming that the values are higher in the high-$\tau$ component than in the low-$\tau$ component in most regions. 
It is worth noting that the radial velocities obtained for the He$\alpha$ emission with the Gaussian modeling (Figure~\ref{fig:zgauss_dop}) are intermediate between those obtained for the high-$\tau$ and low-$\tau$ components in Figure~\ref{fig:doppler_2nei}. 
This is because the He-like emissions of both Si and S are reproduced by a combination of the two components (see Figure~\ref{fig:2nei_all}). 
On the other hand, the Si~\emissiontype{XIV} and S~\emissiontype{XVI} Ly$\alpha$ emissions are predominantly contributed by the high-$\tau$ component in all the regions except SE1. 
Therefore, the radial velocities for the Ly$\alpha$ emissions measured with the Gaussian modeling are comparable to those obtained for the high-$\tau$ component. 
An exceptional result is obtained in the SE1 region, where the electron temperature of the low-$\tau$ component is relatively high ($\sim$\,5\,keV), resulting in the Ly$\alpha$ emission being contributed by both components. 

We note that the absolute abundances (relative to hydrogen) obtained from our analysis are highly uncertain, although their values appear to be well constrained in Table~2. This is because the bremsstrahlung continuum in this energy band could be substantially contributed by heavy elements such as C and O, rather than H. In fact, if we fix the C and O abundances to an extremely large value (i.e., 1000~solar), we still obtain a reasonable fit with similar relative abundances of Si/S. We believe that the observed ejecta component scarcely contains the light elements, since Cas~A is thought to originate from a Type IIb (hydrogen-poor) SN \citep{2008Sci...320.1195K,2011ApJ...732....3R}.

\section{Proper Motion Measurement Using Chandra} \label{sec:4}

In this section, we analyze Chandra archival data to measure the proper motion (i.e., velocity component perpendicular to the line of site) of the IME ejecta, and then determine their three-dimensional velocity by combining it with the XRISM-measured radial velocity. 
Data reduction and analysis procedures follow \citet{2024ApJ...974..245S} with some modifications described below. 
We use Chandra data obtained in 2004 (Obs\,IDs: 4636, 4637, 4639, 5319) and 2019 (Obs\,ID: 19606)\footnote{The data are available at DOI: https://doi.org/10.25574/cdc.309.}$\!$.
Since our analysis in the previous section indicates that the Si~Ly$\alpha$ emission is dominated by the high-$\tau$ component in most regions, we generate images of the 1.95--2.1\,keV band (corresponding to the Si~Ly$\alpha$ emission) for the proper motion measurement.
Unfortunately, we cannot determine the proper motion purely of the low-$\tau$ component, because there is no emission line contributed only by this component (Figure~\ref{fig:2nei_all}).

Following \citet{2023ApJ...951...59S}, we apply the Richardson-Lucy deconvolution \citep{1972JOSA...62...55R,1974AJ.....79..745L} with spatially variant PSF (hereafter RL${\rm{sv}}$). The PSFs are simulated using MARX \citep{2012SPIE.8443E..1AD} assuming a photon energy of 2.05~keV. 
The main difference from the method used by \citet{2024ApJ...974..245S} lies in the handling of continuous data in the RL${\rm{sv}}$ images. Since these images provide real-valued counts rather than integer counts, the Poisson distribution cannot be applied directly. To preserve real-valued information, we employ the continuous Poisson distribution \citep[e.g.,][]{2010ElL....46..631K}, where factorial terms $n!$ in the likelihood function are replaced by the gamma function $\Gamma(n+1)$. 
Statistical uncertainties in the proper motion are calculated based on the $C$-statistics, following the procedure outlined in \citet{2018ApJ...853...46S}. 
Systematic uncertainties are assumed to be $\pm$\,0.5~pixel of the Chandra ACIS detector, corresponding to 0.006\,pc (= 372~km\,s$^{-1}$ $\times$ 15\,yrs) at the distance of 3.4~kpc.

\begin{figure}[t]
    \centering
    \includegraphics[width=0.96\linewidth]{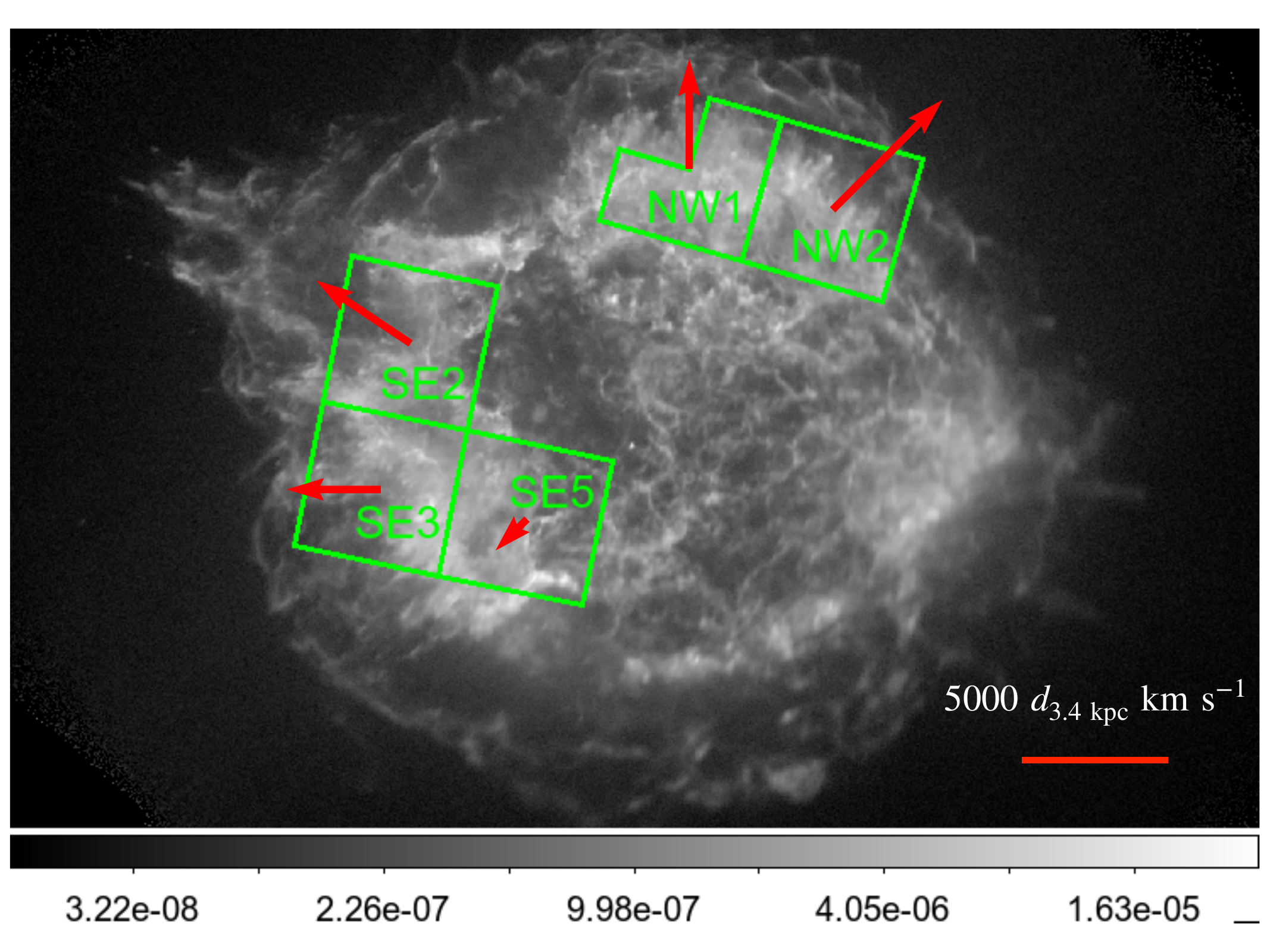}
    \caption{Chandra/ACIS image of Cas~A, overlaid with the regions where the proper motion is measured. 
    The red arrows represent the direction and magnitude of the proper motion measured for the Si~\emissiontype{XIV} Ly$\alpha$ emission (1.95--2.1~keV). 
    }
    \label{fig:proper}
\end{figure}

\begin{table}
  \tbl{Three-dimensional velocity of the IME ejecta}{%
    \begin{tabular}{lcccc}
      \hline
      \hline
        &$v_\mathrm{rad}$&$v_\mathrm{prop}$&$\theta(v_\mathrm{prop})$&$v_{\rm obs}$\\
      Region &(km\,s$^{-1}$)&(km\,s$^{-1}$)&($^\circ$)&(km\,s$^{-1}$)\\ 
      \hline
       SE2&$-1174_{-39}^{+50}$&$3799\pm372$&$56^{+5}_{-6}$&$4000\pm400$ \\
       SE3&$-1192_{-29}^{+26}$&$3161\pm372$&$90^{+4}_{-5}$&$3400\pm400$ \\
       SE5&$-1410_{-33}^{+37}$&$1490\pm372$&$135\pm14$&$2100\pm300$ \\
       NW1&$1665_{-43}^{+45}$&$3688\pm372$&$0\pm4$&$4000\pm400$ \\
       NW2&$1612_{-36}^{+42}$&$5216\pm372$&$315\pm4$&$5500\pm400$ \\
      \hline
    \end{tabular}
  }
  \label{tab:shocked_ejeta_v}
\end{table}

The proper motion measurements are performed for each of the regions where the Resolve spectra are extracted, obtaining the velocity vectors given in Figure~\ref{fig:proper}.
We exclude Regions SE1, SE4, and SE6, because no converging value is obtained for these regions.
In the other regions in both SE and NW rims, the statistical uncertainty in the proper motion is negligibly small compared to the systematic uncertainty.
In Table~\ref{tab:shocked_ejeta_v}, we show the three-dimensional velocities of the IME ejecta in the observer frame ($v_{\rm obs}$), calculated using the radial velocities ($v_{\rm rad}$) 
and the proper motion velocities ($v_{\rm prop}$) measured by XRISM and Chandra, respectively. 
The proper motion velocity is characterized by its magnitude $v_{\rm prop}$ and direction $\theta(v_{\rm prop})$, measured anticlockwise from the celestial north.
We find that the proper motion velocities are about 2--3 times higher than the radial velocities.

\section{Discussion} \label{sec:5}

We have performed Doppler velocity measurements of the IME ejecta in the SNR Cas~A,
utilizing the superb spectral resolution of the XRISM/Resolve. 
Our remarkable finding is that two components of NEI plasmas are required to reproduce the centroid energies and line profiles of the He$\alpha$ and Ly$\alpha$ emissions of the IMEs detected in the 1.75--2.95~keV spectra: the high-$\tau$ component with a higher radial velocity and the low-$\tau$ component with a lower radial velocity. 
The correlation between the velocity and ionization timescale can be explained naturally in the context of the SNR evolution. 
Since the reverse shock generally propagates from the outer layers to the inner layers of an SNR (\citealp{1982ApJ...258..790C}; \citealp{1999ApJS..120..299T}), the ejecta with a higher velocity are shock heated earlier than those with a lower velocity, and thus a higher ionization is achieved in the former \citep{2017hsn..book..875C}.  
In this sense, we can assume that the high-$\tau$ component represents the outermost IME ejecta in this SNR. 
The layered distribution of the ionization degree is also confirmed in Figure~\ref{fig:color_map}. 
The S~Ly$\alpha$/He$\alpha$ flux ratio increases towards the outer regions on the sky plane (which is particularly clear in the SE rim), indicating that more highly ionized ejecta are distributed in the outer layers.

For the high-$\tau$ component, the electron temperature, ionization timescale, and bulk velocity of the shocked ejecta are successfully constrained. 
This allows us to estimate the pre-shock (free expansion) velocity of the outermost IME ejecta and the reverse shock velocity at the time when the ejecta were shock heated as follows. 

In shock-heated plasma in young SNRs where radiative and adiabatic cooling is negligible, the electron temperature evolution is governed mainly by two physical processes: (1) collisionless shock heating, and (2) Coulomb interactions between ions and electrons in postshock plasma. 
In the first process, the relation between the upstream fluid velocity in the shock-rest frame ($v_{u,\rm sh}$) and the downstream temperature ($kT_i$), derived from the Rankine-Hugoniot equations, is given independently for different species $i$ as: 
\begin{equation}\label{eq:RH}
    kT_i=\frac{3}{16} m_i v_{u,\rm sh}^2,
\end{equation}
where $m_i$ is the mass of species $i$. 
Several theoretical and observational studies have, however, suggested that the so-called ``collisionless electron heating'' takes place at the shock front, modifying the  postshock electron temperature to a value higher than the prediction of Equation~\ref{eq:RH} \citep[e.g.,][]{1974ApJ...188..335M,2000ApJS..127..409L,RAKOWSKI20051017,2007ApJ...654L..69G,2023ApJ...949...50R}.
In the case of Tycho's SNR, the electron-to-ion temperature ratio just behind the reverse shock ($\beta = T_{\rm e}/T_{\rm ion}$) is estimated to be $\sim$\,0.01 \citep{Yamaguchi_2014}, indicating that the majority of the thermal energy is possessed by ions in the immediate postshock plasma.

\begin{figure}[t]
    \centering
    \includegraphics[width=0.96\linewidth]{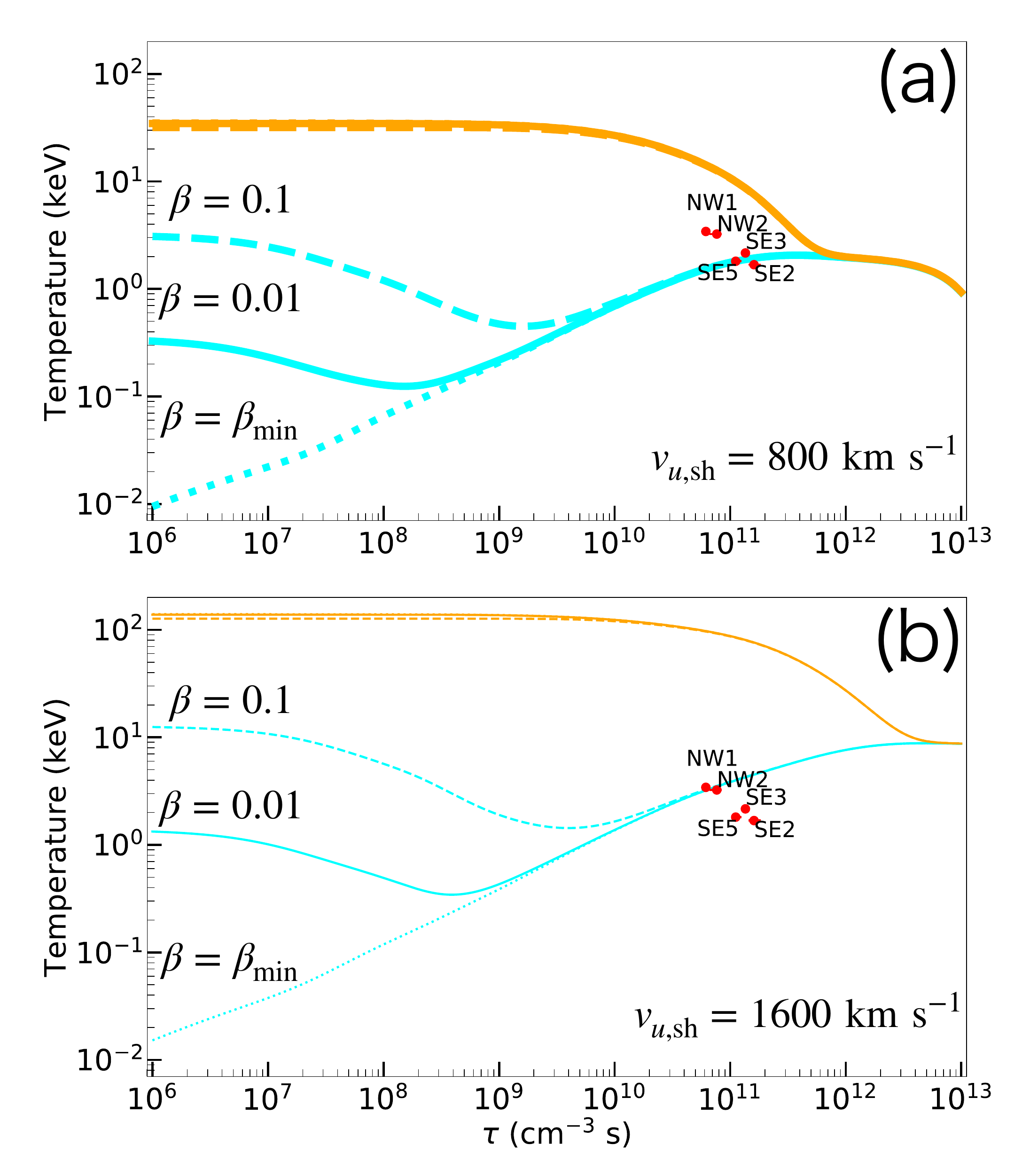}
    \caption{Thermal equilibration between Si temperature (orange) and electron temperature (cyan) via Coulomb collisions in shocked ejecta with pure-Si composition with several values of the initial temperature ratio ($kT_\mathrm{e}/kT_{\mathrm{Si}}$). The upstream bulk velocity is assumed to be 800\,km\,s$^{-1}$ and 1600\,km\,s$^{-1}$ in Panels (a) and (b), respectively. 
    }
    \label{fig:ohshiro}
\end{figure}

In further downstream regions, thermal equilibrium between electrons and ions is gradually achieved through energy exchange via Coulomb collisions. 
The timescale of this process is characterized by a product of the electron density and time after the shock heating, which is identical to the ionization timescale $\tau$. 
Therefore, once the current electron temperature and ionization timescale are measured, we can estimate the initial postshock temperatures by tracing the equilibrium process back to $\tau = 0$, which in turn constrains the upstream fluid velocity using Equation~\ref{eq:RH}. 
In Figure~\ref{fig:ohshiro}, we show the temporal evolution of $kT_{\rm e}$ and $kT_{\rm Si}$ in shocked IME ejecta, calculated using the numerical model of \citet{Ohshiro_2024} for several different values of $\beta$ (= $kT_{\rm e}/kT_{\rm Si}$). 
Here we assume a pure-metal composition, given the Type~IIb SN origin of Cas~A (where the amount of hydrogen in the ejecta is little). 
We find that the plasma properties (i.e., $kT_{\rm e}$ and $\tau$) observed in the SE and NW regions are reasonably reproduced by the models with $v_{u,\rm sh} = 800$~km~s$^{-1}$ (Panel a) and $v_{u,\rm sh} = 1600$~km~s$^{-1}$ (Panel b), respectively, regardless of the assumed $\beta$ value. 
The calculations also provide estimates of current $kT_{\rm Si}$ as $\sim 8$\,keV and $\sim 100$\,keV for the SE and NW regions, respectively. 
The corresponding thermal Doppler broadening of the Si emission is derived to be $\sigma_{\mathrm{th}} \approx
170$~km~s$^{-1}$ for SE and $\sigma_{\mathrm{th}} \approx 590$~km~s$^{-1}$ for NW, using the relation of $\sigma_{\mathrm{th}}^2 = kT_{\rm ion}/m_{\rm ion}$. These values are significantly lower than the $\sigma_v$ values in Table~\ref{tab:2nei_res}, suggesting that the observed line broadening is dominated by the dispersion of the bulk velocity rather than by the thermal Doppler effect.

\begin{table}
  \tbl{$V_{\rm RS}$ and $V_{\rm free}$ (see text for definitions) estimated for each region.}{%
    \begin{tabular}{lccc}
      \hline
      \hline
       & $V_{\rm RS}$ & $V_{\rm free}$ \\
      Region & (km\,s$^{-1}$)& (km\,s$^{-1}$)\\ 
      \hline
       SE2 & $3800\pm400$ & $4600\pm400$ \\
       SE3 & $3200\pm400$ & $4000\pm400$ \\
       SE5 & $1900\pm300$ & $2700\pm300$ \\
       NW1 & $3600\pm400$ & $5200\pm400$ \\
       NW2 & $5100\pm400$ & $6700\pm400$ \\
      \hline
    \end{tabular}
  }
  \label{tab:v_cal}
\end{table}

The Rankine-Hugoniot equations, which have led to Equation~\ref{eq:RH}, also predict the relation between the upstream and downstream bulk velocities in the shock rest frame as 
\begin{equation}
    v_{d,\mathrm{sh}}=\frac{1}{4}v_{u,\mathrm{sh}}.
    \label{eq:RH_v}
\end{equation}
Therefore, the velocity of the shocked ejecta in the observer frame is given as: 
\begin{equation}
    v_{\mathrm{obs}} = V_{\mathrm{RS}} + v_{d,\mathrm{sh}}
    = V_{\mathrm{RS}} + \frac{1}{4}v_{u,\mathrm{sh}},
    \label{eq:v_obs}
\end{equation}

where $V_{\mathrm{RS}}$ is the reverse shock velocity in the observer frame at the time when the ejecta were shock heated. 
Since $v_{\mathrm{obs}}$ and $v_{u,\mathrm{sh}}$ are already determined (in Section~4 and in the previous paragraph, respectively), $V_{\mathrm{RS}}$ can be calculated using the relation converted from Equation~\ref{eq:v_obs}, 
\begin{equation}
    V_{\mathrm{RS}} = 
    v_{\mathrm{obs}} - \frac{1}{4}v_{u,\mathrm{sh}}.
    \label{eq:v_RS}
\end{equation}
We can also estimate the free expansion velocity of the preshock ejecta using the relation 
\begin{equation}
  V_{\mathrm{free}} = v_{u,\mathrm{sh}} + V_{\mathrm{RS}}. 
  \label{eq:v_free}    
\end{equation}
The results for each region are given in Table~\ref{tab:v_cal}. 
The positive values of $V_{\mathrm{RS}}$ indicate that the outermost IME ejecta were heated in the early stage of the SNR evolution, when the reverse shock was propagating outward.
We find that the free expansion velocity of the IME ejecta ranges from 2400 to 7100~km\,s$^{-1}$, substantially higher than typical values predicted for a core-collapse SN of a 15$M_\odot$ progenitor with a hydrogen envelope \citep[$\sim$\,1000~km~s$^{-1}$: e.g.,][]{1995ApJS..101..181W}.
However, pieces of observational evidence indicate that Cas~A has lost most of its hydrogen envelope prior to the explosion \citep[e.g.,][]{1991ApJ...371..621F,2008ApJ...681L..81R}, allowing the IME to be ejected at a higher velocity.
In fact, \citet{2017ApJ...842...13W} performed 3D hydrodynamical SN simulations of a progenitor stripped of most of its hydrogen and predicted that the free expansion velocity of the Si ejecta would be in the range of 2000 to 7000~km\,s$^{-1}$, 
consistent with our estimate.

\begin{figure}[t]
    \centering
    \includegraphics[width=0.96\linewidth]{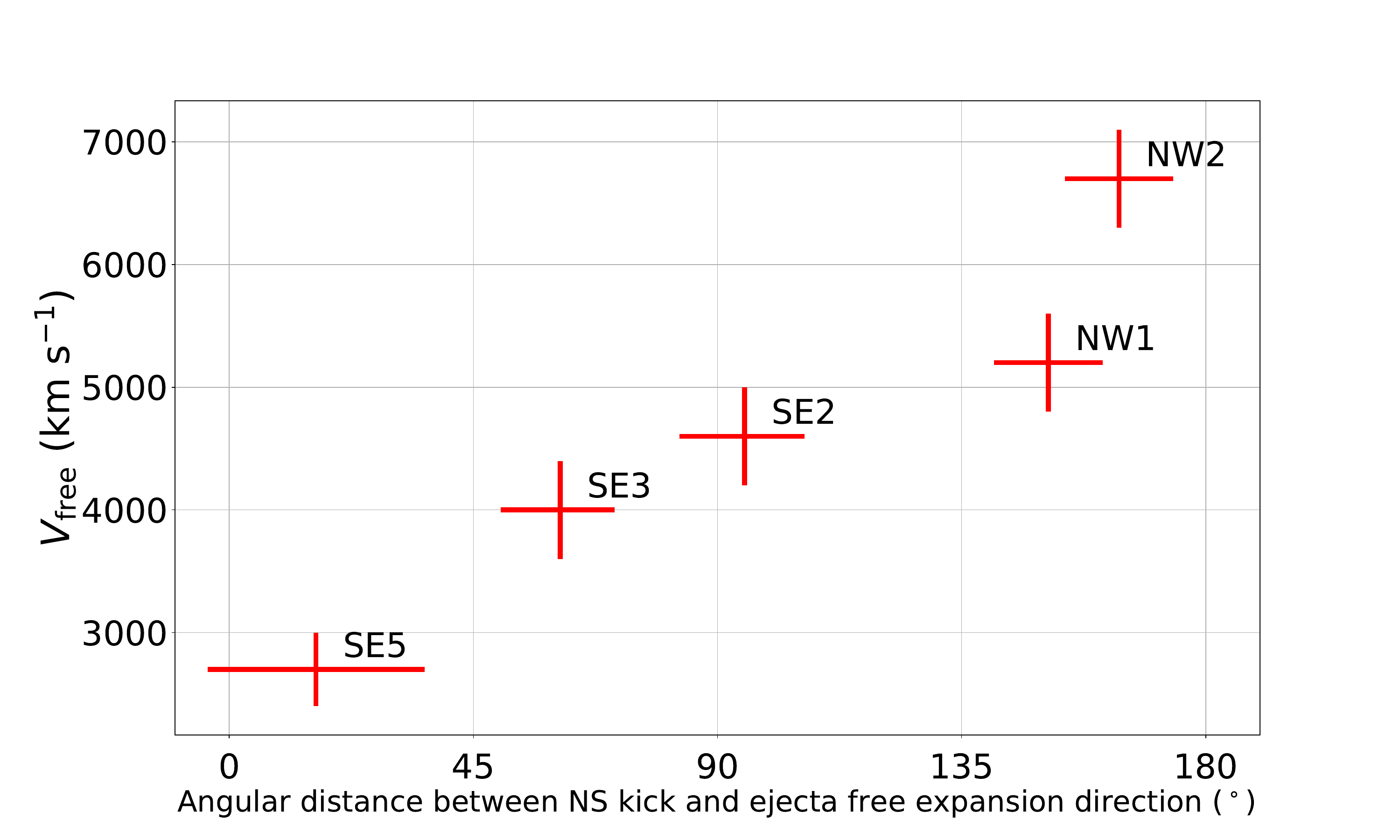}
    \caption{Free expansion velocity of the IME ejecta as a function of angular distance between the directions of the NS motion and ejecta's free expansion velocity.
    }
    \label{fig:nk_dire}
\end{figure}

We have also revealed a substantial asymmetry in the distribution of the free expansion velocity. 
In particular, the highest velocity is observed in the NW2 region, while the lowest velocity is found in the SE5 region. Notably, these directions are almost exactly opposite to and aligned with the proper motion direction of the NS in this SNR, $\sim$151$^\circ$, measured anticlockwise from the celestial north \citep{Katsuda_2018,2024ApJ...962...82H}. 
Figure~\ref{fig:nk_dire} shows the relation between $V_{\mathrm{free}}$ and the angular distance between the directions of the free expansion and NS motion (where we implicitly assume that $\theta (v_{\rm prop})$ in Table~4 is unchanged from the free expansion direction). 
We find a clear correlation between the two quantities, implying that the IME ejecta were expelled more strongly to the direction opposite to the NS motion.
This is another piece of evidence for physical association between asymmetric SN explosion and NS kick, following e.g., \citet{Grefenstette_2014}, \citet{Katsuda_2018}, and \citet{2024ApJ...962...82H}.
Furthermore, the fact that the lowest free expansion velocity is found in the same direction as the NS motion favors the so-called gravitational tug-boat mechanism \citep{2013A&A...552A.126W,2017ApJ...837...84J}.
In this scenario, an asymmetric mass ejection during the explosion creates a prolonged gravitational pull from the slowly expanding ejecta, accelerating the NS in this direction.
Note that our work has provided the first evidence for this relation with the three-dimensional velocity distribution of the IME ejecta, complement to the previous studies, where the spatial distribution of $^{44}$Ti or IME were investigated.

\section{Conclusions} \label{sec:6}

We have presented spatially-resolved, high-resolution spectroscopy of the IME ejecta in the SNR Cas~A, using the Resolve instrument aboard XRISM. 
The 1.75--2.95~keV spectra extracted from the SE and NW regions are well modeled with two-component plasmas with different thermal parameters ($kT_{\rm e}$ and $\tau$) and radial velocity; the presence of the two components is suggested by the fact that the observed centroid energies of the He$\alpha$ and Ly$\alpha$ emissions of Si and S cannot be explained by a single velocity shift value. 
The high-$\tau$ component is found to have a higher radial velocity than the low-$\tau$ component. 
This indicates that more highly-ionized ejecta are distributed in the outer layers, consistent with a scenario that the SNR reverse shock heats the ejecta in the outer layers first. 
Using the Chandra archival data, we have also performed the proper motion measurement of the Si~\emissiontype{XIV}~Ly$\alpha$ emission (which represents the outermost IME ejecta), and have successfully reconstructed their three-dimensional bulk velocity. 
The pre-shock (free expansion) velocity of the outermost IME ejecta is estimated to be 2400--7100~km~s$^{-1}$, based on the thermal properties as well as the bulk velocity of the shocked ejecta. 
These values are consistent with theoretical predictions for a Type IIb SN, where the majority of progenitor's hydrogen envelope has been stripped before the explosion.
We have revealed that the highest free expansion velocity is achieved in the same direction as the NS motion, suggesting a physical relation between the asymmetric SN explosion and the NS kick.

\bibliography{references}{}
\bibliographystyle{apj}


\section*{Acknowledgments}
The authors thank the XRISM Science Team members, especially Paul Plucinsky and Toshiki Sato for their leadership in the activities of the Cas~A target team.
We also express our gratitude to Daiki Miura, Yuki Amano, Hiromasa Suzuki, Hirofumi Noda, and Richard Mushotzky for their helpful advice on data analysis and discussion.
\section*{Appendix}
\begin{figure*}
    \centering
    \includegraphics[width=0.96\linewidth]{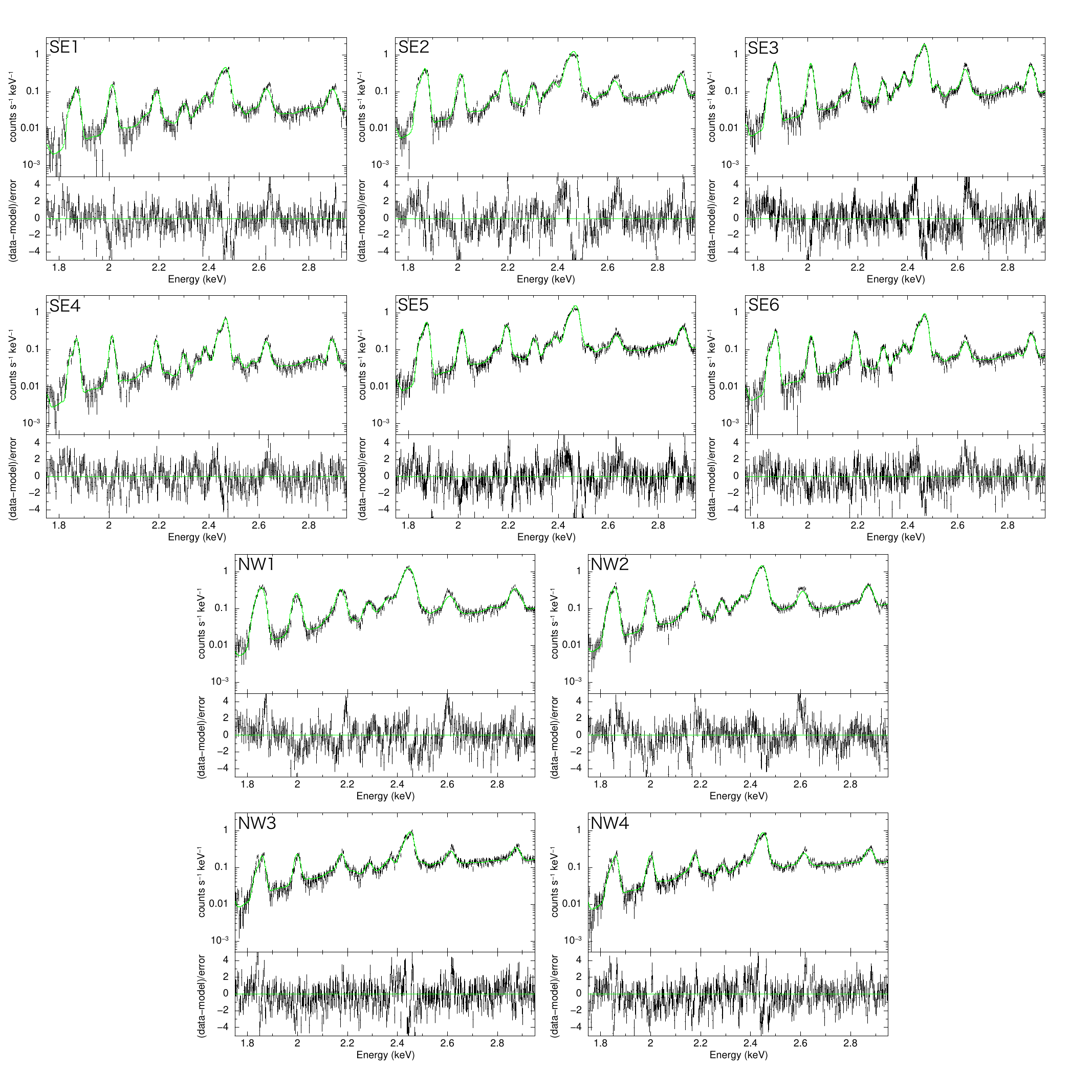}
    \caption{The Resolve spectra in the 1.75--2.95~keV band from all regions, fit with the single non-equilibrium ionization (NEI) model (green).
    }
    \label{fig_1nei_AL}
\end{figure*}

\begin{figure*}
    \centering
    \includegraphics[width=0.96\linewidth]{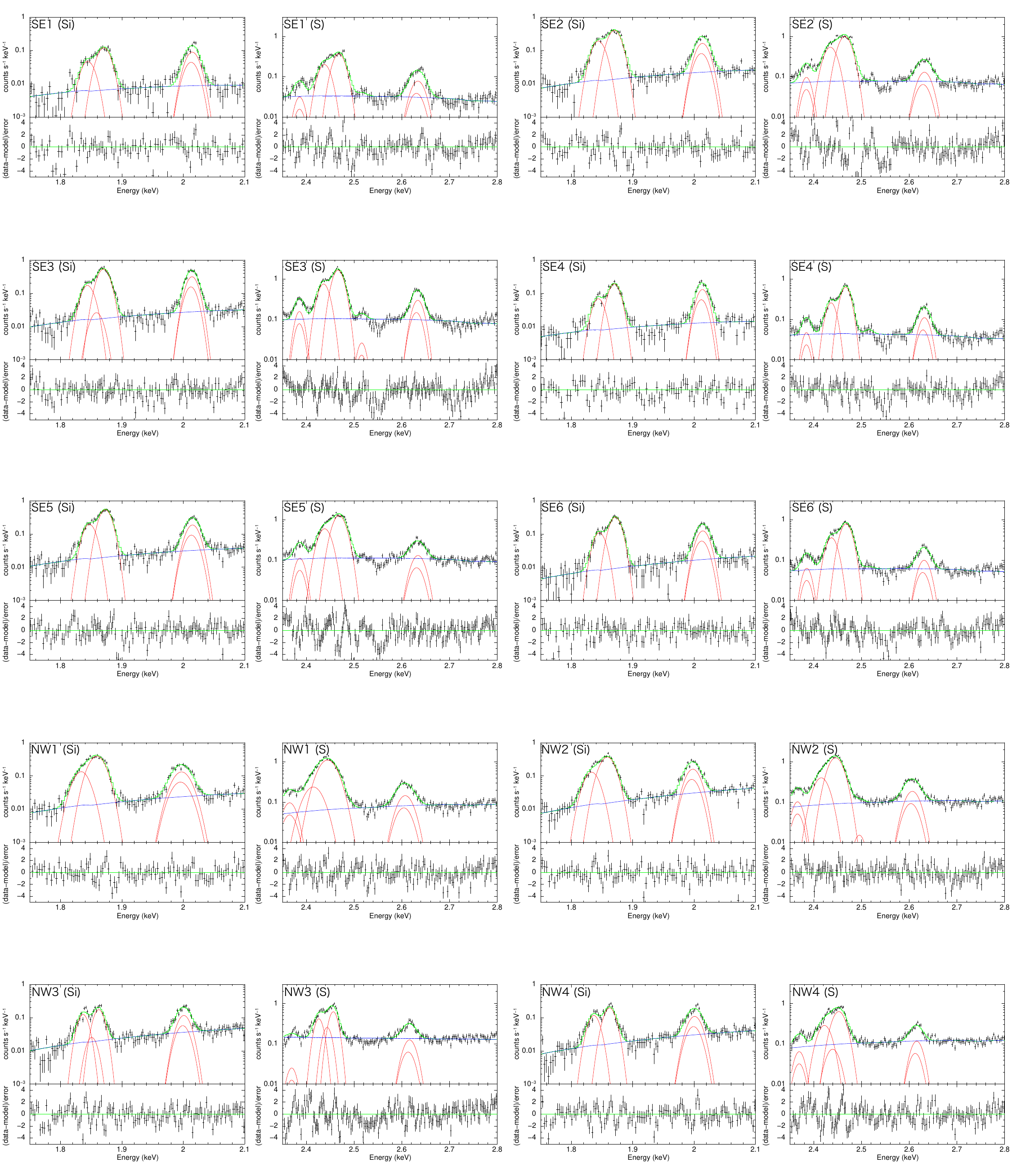}
    \caption{The Resolve spectrum in the 1.75--2.1~keV and 2.35--2.8~keV bands, corresponding to the Si and S bands fitted with the empirical model (red). The contributions of the Gaussian functions bremsstrahlung continuum component are indicated as red and blue, respectively.
    }
    \label{fig:zgauss_all}
\end{figure*}

Table~\ref{tab:1nei} lists the best-fit values for the single component (\texttt{bvrnei}) model and $C$-stat/dof for all regions.
Figure~\ref{fig_1nei_AL} shows the spectra and best-fit model of all regions obtained using the single component (\texttt{bvrnei}) model.
Figure~\ref{fig:zgauss_all} displays the spectra and best-fit models of all regions obtained using Gaussian models.

\begin{table*}
  \tbl{Best-fit parameters of the single non-equilibrium ionization (NEI) model fit in the 1.75--2.95~keV band.}{%
    \begin{tabular}{lcccccccc}
\hline
\hline
       & $kT_\mathrm{e}$                    & Si abundance           & S abundance            & $\tau$                  & $v_{\rm shift}$                     & $\sigma_v$         &                        &              \\
Region & (keV)                     & (solar)                & (solar)                & ($10^{11}$ s~cm$^{-3}$) &(km~s$^{-1}$)       & (km~s$^{-1}$)      & normalization          & $C$-stat/dof \\ \hline
SE1&$1.37^{+0.05}_{-0.02}$&$5.4^{+0.2}_{-0.4}$&$4.0^{+0.1}_{-0.2}$&$2.11^{+0.24}_{-0.22}$&$-953^{+30}_{-35}$&$1585^{+26}_{-36}$&$1.74^{+0.12}_{-0.07}$&$757.96/343$ \\
SE2&$1.21^{+0.02}_{-0.03}$&$5.2\pm0.1$&$3.66^{+0.08}_{-0.07}$&$1.61^{+0.08}_{-0.04}$&$-697\pm18$&$1571^{+16}_{-17}$&$1.97^{+0.11}_{-0.07}$&$1468.67/380$ \\
SE3&$1.373^{+0.005}_{-0.002}$&$6.1\pm0.1$&$4.50^{+0.08}_{-0.09}$&$1.66\pm0.06$&$-913\pm12$&$1312\pm10$&$1.48^{+0.04}_{-0.01}$&$1354.23/479$ \\
SE4&$1.40\pm0.04$&$4.7^{+0.1}_{-0.2}$&$3.4\pm0.1$&$1.77^{+0.11}_{-0.08}$&$-906\pm19$&$1205\pm17$&$1.72^{+0.04}_{-0.08}$&$626.68/352$ \\
SE5&$1.31^{+0.03}_{-0.02}$&$4.22^{+0.09}_{-0.08}$&$2.95\pm0.06$&$1.21^{+0.07}_{-0.08}$&$-1188^{+15}_{-16}$&$1496\pm15$&$2.19^{+0.08}_{-0.07}$&$1257.52/485$ \\
SE6&$1.41^{+0.04}_{-0.03}$&$4.2\pm0.1$&$2.97\pm0.07$&$1.19\pm0.07$&$-1026^{+18}_{-19}$&$1319^{+16}_{-8}$&$2.09^{+0.08}_{-0.09}$&$743.25/421$ \\
NW1&$1.64^{+0.05}_{-0.02}$&$5.99^{+0.15}_{-0.09}$&$4.24^{+0.13}_{-0.09}$&$0.89^{+0.04}_{-0.07}$&$1444^{+27}_{-20}$&$2049^{+20}_{-19}$&$1.45^{+0.05}_{-0.07}$&$924.30/383$ \\
NW2&$1.64^{+0.02}_{-0.05}$&$4.02^{+0.08}_{-0.10}$&$3.12^{+0.07}_{-0.06}$&$1.01^{+0.04}_{-0.07}$&$1504^{+19}_{-20}$&$1726\pm18$&$2.17^{+0.10}_{-0.05}$&$876.90/386$ \\
NW3&$1.71^{+0.03}_{-0.06}$&$1.53^{+0.04}_{-0.03}$&$1.29^{+0.03}_{-0.03}$&$1.13^{+0.09}_{-0.05}$&$1028^{+28}_{-33}$&$1564^{+28}_{-32}$&$3.88^{+0.18}_{-0.07}$&$1010.66/473$ \\
NW4&$1.70^{+0.04}_{-0.08}$&$1.79^{+0.06}_{-0.06}$&$1.48^{+0.03}_{-0.04}$&$1.08^{+0.10}_{-0.05}$&$1005^{+31}_{-36}$&$1615^{+29}_{-35}$&$3.45^{+0.21}_{-0.10}$&$899.39/439$ \\
\hline
\end{tabular}
  }
  \label{tab:1nei}
\end{table*}

\end{document}